\documentclass[]{aa}
\usepackage{lscape}
\usepackage{amsmath}
\usepackage{amssymb}
\usepackage{natbib}
\usepackage{graphicx}
\usepackage{txfonts}
\usepackage[english]{babel}
\usepackage{amsfonts}

\newcommand{\parameters}{
\begin{table}
\caption{Characterizing parameters of HD\,142527. }
\label{tab:parameters} 
\centering
\begin{tabular}{l c} 
\hline\hline\\[-9pt]   
Parameter                 & Value         \\  
\hline\\[-9pt]                      
Right Ascension           & 15$^{\rm h}$ 56$^{\rm m}$ 41\,$\fs$89    \\
Declination               & -42$^\circ$ 19$\arcmin$ 23$\farcs$5    \\
Spectral Type             & F6IIIe        \\
$T_{\rm eff}$\,[K]        & 6250          \\
Distance\,[pc]            & 145$\pm$15    \\
$A_{\rm V}$               & 0.60$\pm$0.05 \\
Luminosity\,[L$_{\odot}$] & 20$\pm$2      \\
Radius\,[R$_\odot$]       & 3.8$\pm$0.3   \\
Mass\,[M$_{\odot}$]       & 2.2$\pm$0.3   \\
log Age\,[yr]                   & 6.7$\pm$0.4 \\
Group                     & Ia            \\
\hline
\end{tabular}
\tablefoot{\ch{We \ct{obtained new stellar parameters from our disk modeling effort,} see the text for details.}}
\end{table}
}

\newcommand{\mmobs}{
\begin{table}
\caption{ATCA fluxes and positions. }
\label{tab:mmobs}
\centering
\begin{tabular}{lccc} 
\hline\hline\\[-9pt]
Name     &   Peak flux\,[mJy] & $R.A.$                              & $\delta$    \\
\hline\\[-9pt]
Uranus   &   7.04           & -                                     & -            \\
IRSV\,1540 &  10.8            & 15$^{\rm h}$ 44$^{\rm m}$ 39\,$\fs$90  & -54$^\circ$ 23$\arcmin$ 04$\farcs$4 \\
HD\,142527 &  43.1$\pm$5.4    & 15$^{\rm h}$ 56$^{\rm m}$ 41\,$\fs$87  &  -42$^\circ$ 19$\arcmin$ 23$\farcs$5  \\
\hline
\end{tabular}
\tablefoot{The reported measurements of the calibrators and 
the target are obtained at 3.5\,mm. For the phase-calibrator the flux was derived from 
imaging the six channels centered on the maser line. }
\end{table}
}

\newcommand{\logima}{
\begin{table}
\caption{Log of the VISIR imaging. }
\label{tab:logima}
\centering
\scriptsize
\begin{tabular}{l c c c c c c c} 
\hline\hline\\[-7pt]  
Target      & Filter & $\lambda_c$ & Chop           & Time  & Airmass & Seeing      & Int.time \\
                &          & [$\mu$m]       & [$\arcsec$]  & [h:m] &         & [$\arcsec$] & [s]      \\
\hline\\[-7pt]                          
HD\,139127  &    SiC & 11.85 &  9 & 05:07   &   1.07 &    0.83 &  110 \\
HD\,142527  &    SiC & 11.85 & 10 & 05:18   &   1.07 &    0.83 &  354 \\
HD\,186791  &    SiC & 11.85 &  9 & 10:07   &   1.22 &    0.77 &  110 \\
HD\,142527  &     Q2 & 18.72 &  8 & 02:21   &   1.05 &    0.82 &  680 \\
HD\,142527  &     Q2 & 18.72 &  8 & 02:37   &   1.06 &    0.75 &  680 \\
HD\,139127  &     Q2 & 18.72 &  9 & 02:57   &   1.09 &    0.72 &  338 \\
\hline
\end{tabular}
\tablefoot{The entries are ordered according to observation time. The columns from left to right give the target name, the 
imaging filter, \ch{its central wavelength,} the chopper throw, the time of observation, the airmass and 
optical seeing, and the \ch{effective} integration time. }
\normalsize
\end{table}
}

\newcommand{\logmr}{
\begin{table}
\caption{Log of the VISIR spectroscopy.  }
\label{tab:logmr}
\centering
\scriptsize
\begin{tabular}{l c c r c c c r}
\hline\hline\\[-7pt]
Target     & R   & $\lambda_c$   & Orient.      & Time  & Airmass & Seeing      & Int.time \\
               &      & [$\mu$m]         & [$^{\circ}$]  & [h:m] &         & [$\arcsec$] & [s]      \\
\hline\\[-7pt]
HD\,142527     & LR  &    8.8 &     0 & 05:58   &   1.05 &   0.76 &  644  \\
HD\,142527     & LR  &    9.8 &     0 & 06:13   &   1.05 &   0.76 &  644  \\
HD\,139127     & LR  &    8.8 &     0 & 06:38   &   1.07 &   0.67 &  91  \\
HD\,139127     & LR  &    9.8 &     0 & 06:42   &   1.07 &   0.60 &  91  \\
HD\,139127     & LR  &   11.4 &     0 & 06:45   &   1.07 &   0.61 &  91  \\
HD\,142527     & LR  &   11.4 &     0 & 07:01   &   1.07 &   0.67 &  644  \\
HD\,142527     & LR  &   12.4 &     0 & 07:17   &   1.09 &   0.67 &  517  \\
HD\,139127     & LR  &   12.4 &     0 & 07:40   &   1.14 &   0.73 &  97  \\
\hline\\[-7pt]
HD\,142527     & MR  & 18.18 &   113 & 03:10   &   1.11 &   1.19 &  1846  \\
HD\,142527     & MR  & 17.82 &   113 & 03:52   &   1.18 &   1.40 &  1846  \\
HD\,148478     & MR  & 18.18 &     0 & 05:17   &   1.25 &   0.86 &  159  \\
HD\,148478     & MR  & 17.82 &     0 & 05:23   &   1.27 &   0.83 &  159  \\
\hline
\end{tabular}
\tablefoot{The entries are ordered according to observation time. The columns from 
left to right give the target name, \ch{the spectral resolution (low/medium), the
central wavelength}, the orientation of the slit
(rotating from north to west), the time of observation, the airmass and
optical seeing, and the \ch{effective} integration time.  }
\normalsize
\end{table}
}

\newcommand{\phot}{
\begin{table}
\caption{Photometric fluxes.}
\label{tab:phot}
\centering
\begin{tabular}{l c c c}
\hline\hline\\[-9pt]
Band  & $\lambda$\,[$\mu$m] & F\,[Jy] & Reference \\
\hline\\[-9pt]
Johnson U  & 0.36   & 0.34$\pm$0.01     & M98 \\
Johnson B  & 0.44   & 1.12$\pm$0.01     & M98 \\
Johnson V  & 0.55   & 1.84$\pm$0.02     & M98 \\
Near-IR J    & 1.23   & 3.8$\pm$0.1         & M98 \\
Near-IR H   & 1.65   & 4.6$\pm$0.1         & M98 \\
Near-IR K   & 2.22   & 5.6$\pm$0.1         & M98 \\
Near-IR L    & 3.77   & 7.4$\pm$0.3         & M98 \\
Near-IR M   & 4.78   &  6.7$\pm$0.3        & M98 \\
SiC              & 11.85  & 8.8$\pm$0.6         & V10 \\
IRAS 12      & 11.80  &  10$\pm$4            & B88 \\
Q2              & 18.72  &  14.6$\pm$1.5      & V10 \\
IRAS 25     &  24.38  &   21$\pm$5           & B88 \\
IRAS 60     & 58.61   &   105$\pm$12       & B88 \\
IRAS 100   & 100.9   &   84.7$\pm$16     & B88 \\
CSO  0.8    & 800    & 4.2$\pm$0.5          & W95 \\
SEST 1.2    & 1200  & 1.12$\pm$0.02     & V10 \\
CSO 1.3    & 1300  & 1.19$\pm$0.03      & W95 \\
ATCA 3.5    & 3476  & 0.047$\pm$0.006  & V10 \\ 
\hline\\[-9pt]
\end{tabular}
\tablebib{B88 = \citealt{1988iras....1.....B}; M98 = \citealt{1998A&A...331..211M}; 
V10 = this paper;  W95 = \citealt{1995Ap&SS.224..389W}. }
\end{table}
}

\newcommand{\modelfit}{
\begin{table}
\caption{The parameters \ch{derived with} the best fit model.}
\label{tab:modelfit}
\centering
\begin{tabular}{lccc}
\hline\hline\\[-9pt]
Parameter   & Inner disk & Halo & Outer disk \\
\hline\\[-9pt]
Inner radius ($R_{\rm in}$\,[AU])  & $0.3$           & $0.3$       & $130$ \\
Outer radius ($R_{\rm out}$\,[AU]) & $30$            & $30$        & $200$ \\
Inclination ($i$)                  & $20^\circ$      & -  & $20^\circ$ \\
Position Angle ($PA$)          	   & $160^\circ$     & - & $160^\circ$ \\
Sedimentation par. ($\Psi$)   	   & $0.4$	     & -           & $0.9$ \\
Power law exponent	($p$)      & -1              & -1          & -1     \\
Dust mass ($M$\,[M$_{\odot}$])	   & 2.5$\cdot$10$^{-9}$  & 1.3$\cdot$10$^{-10}$ & 1.0$\cdot$10$^{-3}$ \\
Silicate abundance [\%]            & 63              & 63          & 34  \\
Carbon abundance [\%]              & 37              & 37          & 20   \\
Ice abundance [\%]                 & -               & -           & 45  \\
Grain size [$\mu$m]                & 0.1,1.5$^*$     & 0.1,1.5$^*$ & 2.5 \\
Crystallinity [\%]                 & 23              & 23        & 0      \\
\hline\\[-9pt]
\end{tabular}
\tablefoot{\ch{Abundances are given as mass fractions.} (*) Following \cite{2005A&A...437..189V}.}
\end{table}
}

\newcommand{\ATCA}{
\begin{figure}[t!]
\centering
\includegraphics[width=0.87\columnwidth]{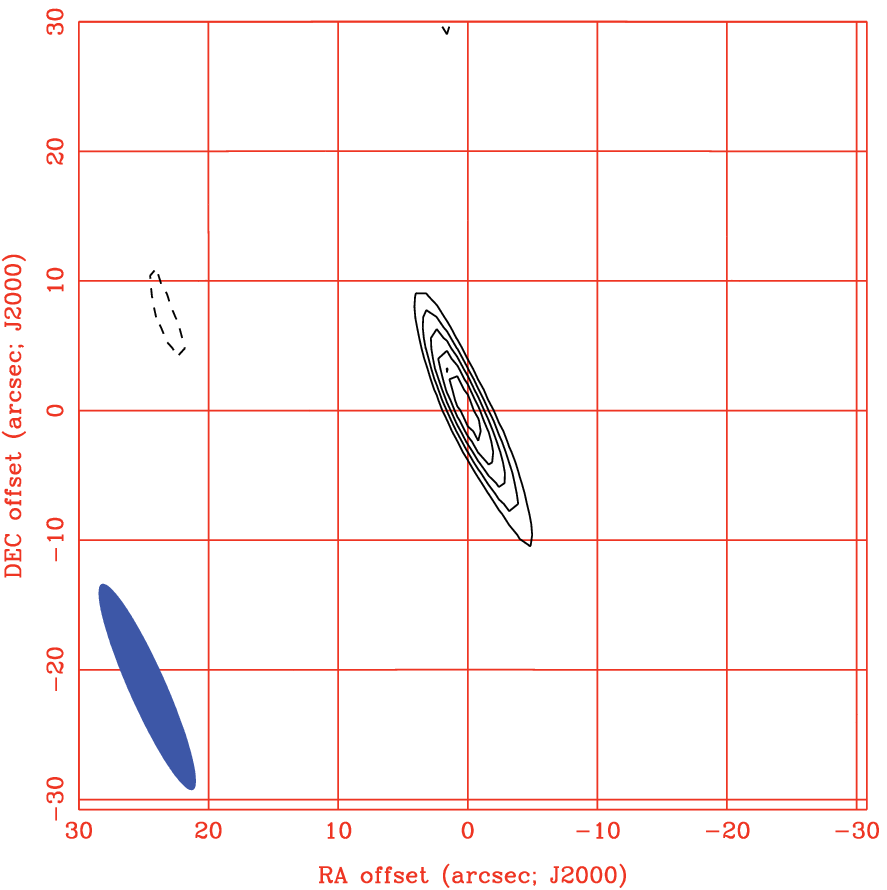}
\caption{\ch{ATCA 3.5\,mm image centered on the peak of the emission
of HD\,142527. } The displayed contour levels are -16.2 \ch{(dotted line)},16.2, 21.6, 27.0,
32.4, 37.8 mJy/beam. The resolution of the image is 16.2$\arcsec$ by
2.9$\arcsec$ at a $PA$ of 24$^{\circ}$. The beam size is shown in the
bottom left. The source appears to be slightly resolved. }
\label{fig:ATCA}
\end{figure}
}

\newcommand{\spitzer}{
\begin{figure}[t!]
\includegraphics[width=\columnwidth]{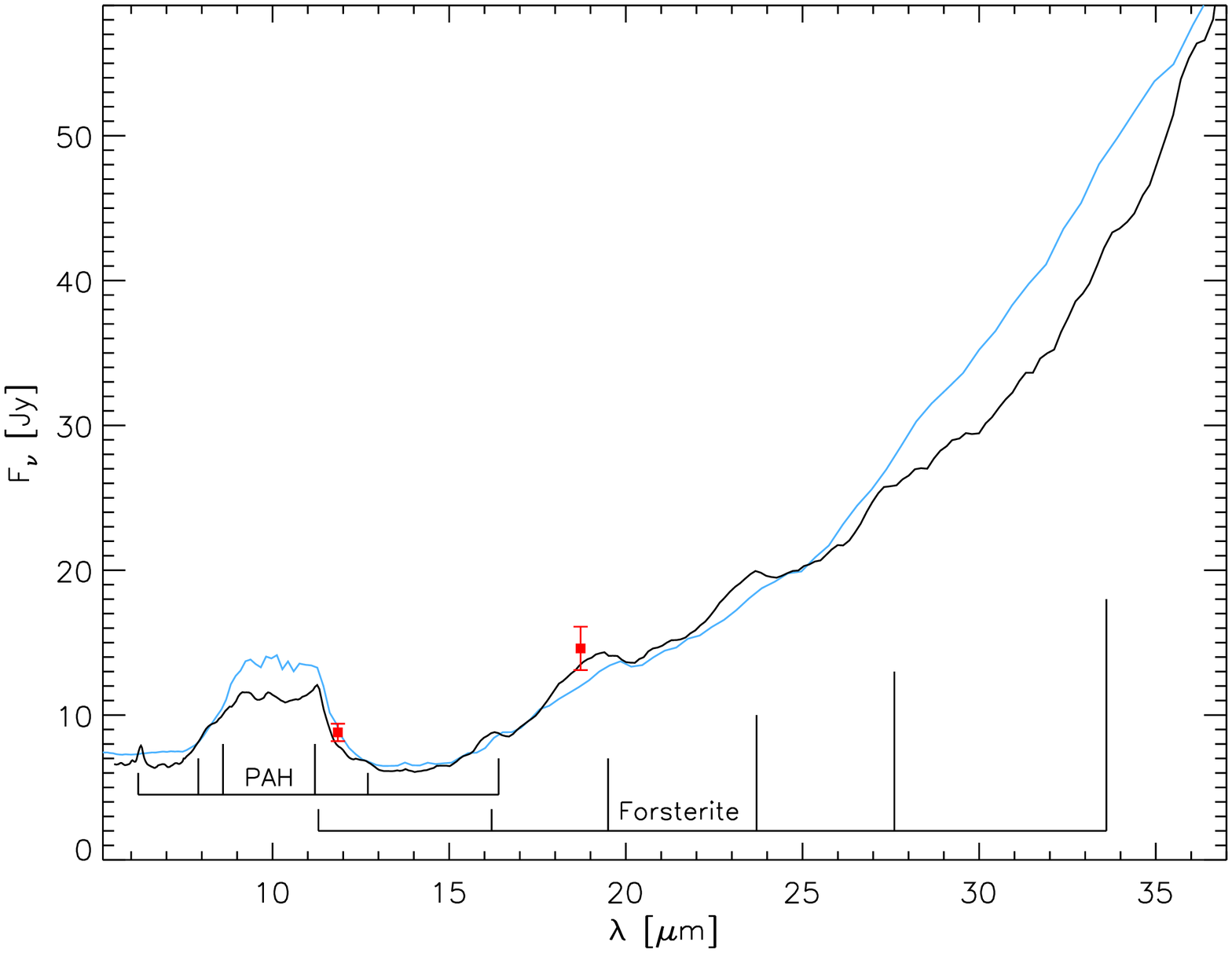}
\caption{Spitzer-IRS LR spectrum of HD\,142527 (SNR $\approx$ 400). 
Indicated are the wavelengths of the PAH bands (at 6.2, 7.9, 8.6,
11.2, 12.7, and 16.4~$\mu$m), and of the forsterite bands (at 11.3,
16.2, 19.5, \ct{23.7, 27.6 and 33.6}~$\mu$m). The photometric points (red) are
from our VISIR imaging. In blue we overplotted the modeled spectrum.}
\label{fig:spitzer}
\end{figure}
}

\newcommand{\slitpos}{
\begin{figure}[t!]
\includegraphics[width=\columnwidth]{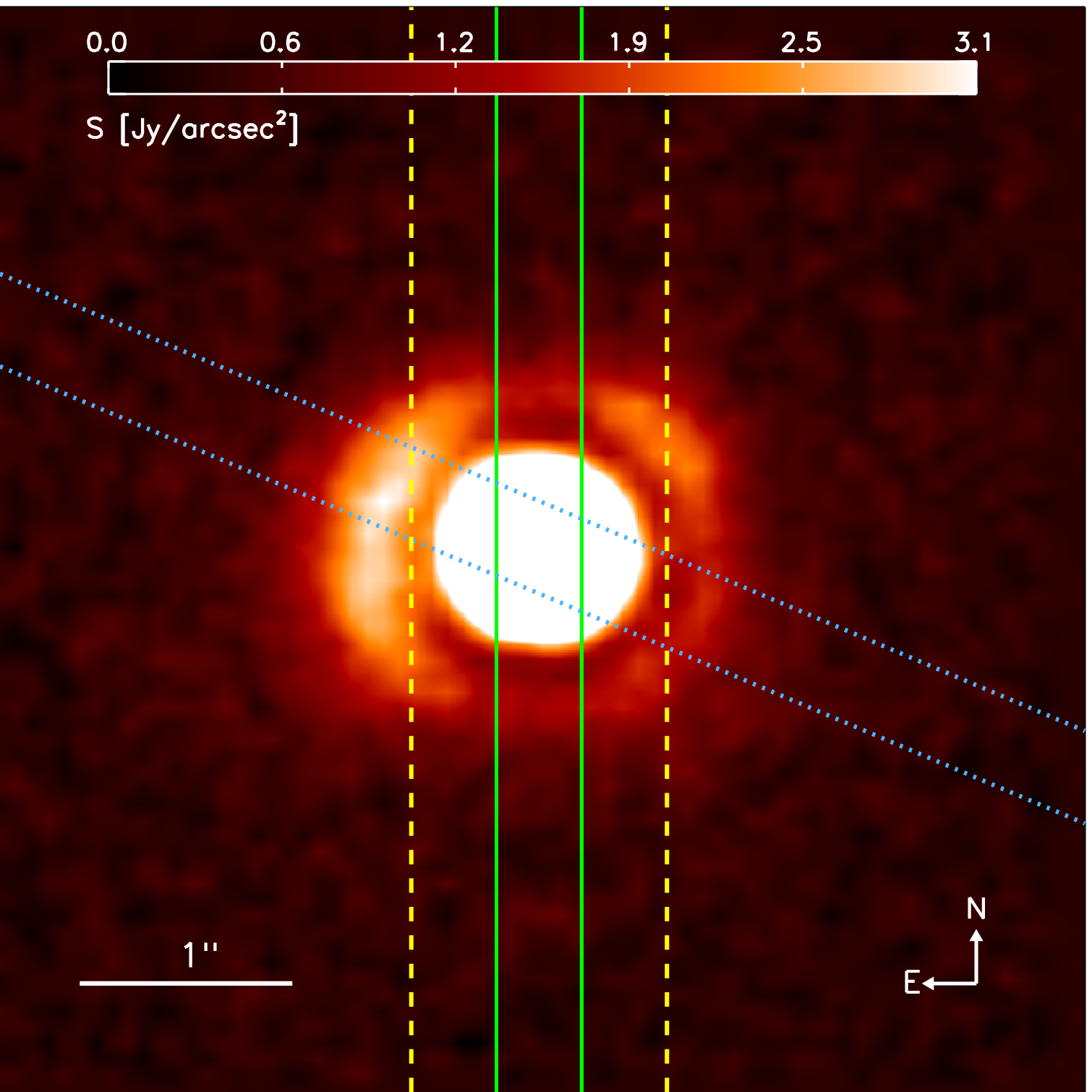}
\caption{Observed 18.72~$\mu$m VISIR image. The color bar shows 
the surface brightness in Jy/arcsec$^2$. Overplotted are the positions
of the VISIR and TIMMI2 spectroscopic slits. The VISIR N-band slit
position is plotted with the green straight line, the VISIR Q-band
slit position is plotted with the blue dotted line and the TIMMI2
N-band spectrum is plotted with the yellow dashed line. The Spitzer
slit encompasses the entire image. }
\label{fig:slitpos}
\end{figure}
}

\newcommand{\contour}{
\begin{figure}[t!]
\includegraphics[width=\columnwidth]{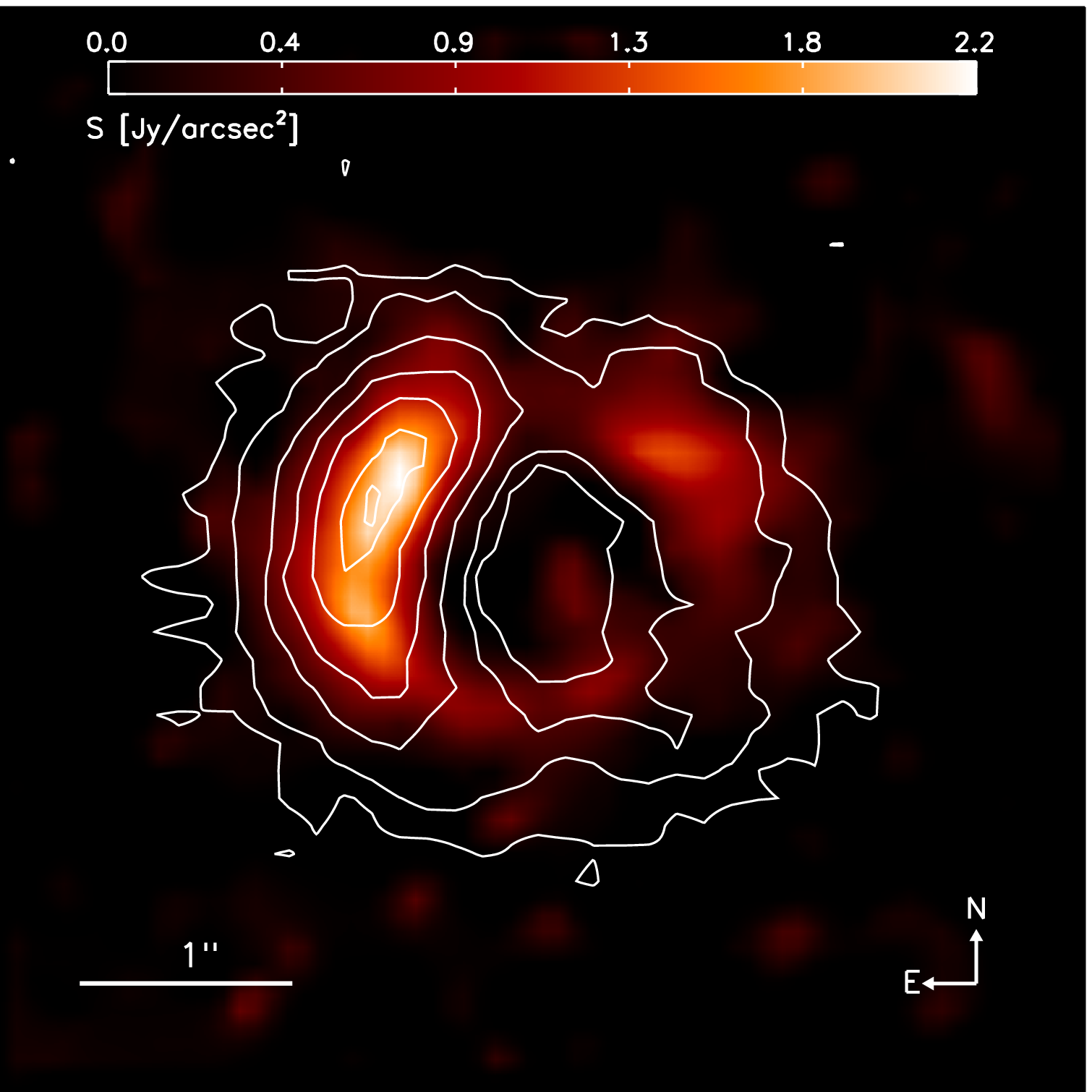}
\caption{Central component-subtracted VISIR 18.72\,$\mu$m
image of HD\,142527. The color bar shows the surface brightness with a
cut-off at 3.1\,Jy/arcsec$^2$. The overplotted contours from the
24.5\,$\mu$m Subaru image are at $0.5,1.0,2.0,3.0,4.0,5.0$, and
$5.3$\,Jy/arcsec$^2$.}
\label{fig:contour}
\end{figure}
}

\newcommand{\iprofile}{
\begin{figure}[t!]
\includegraphics[width=\columnwidth]{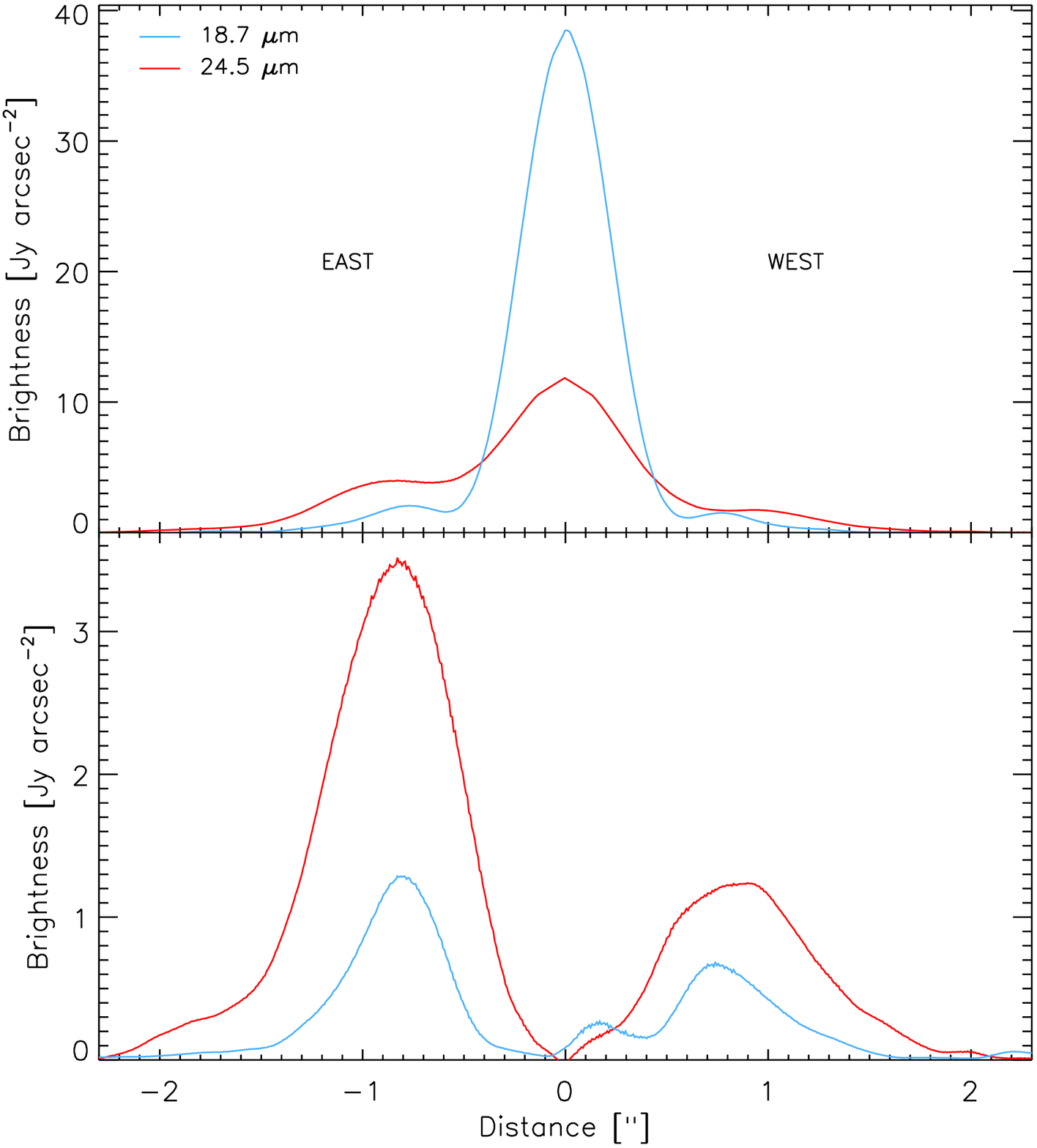}
\caption{Mean radial surface brightness profiles of the east 
($PA$ = 0-180$^\circ$) and west ($PA$ = 180-360$^\circ$) sides of the
VISIR 18.72\,$\mu$m image (blue) and the COMICS 24.5\,$\mu$m image
(red). The top panel shows the profile of the observed images and the
bottom panel shows the profile after central component subtraction. }
\label{fig:iprofile}
\end{figure}
}

\newcommand{\innerdisk}{
\begin{figure}[t!]
\centering
\includegraphics[width=\columnwidth]{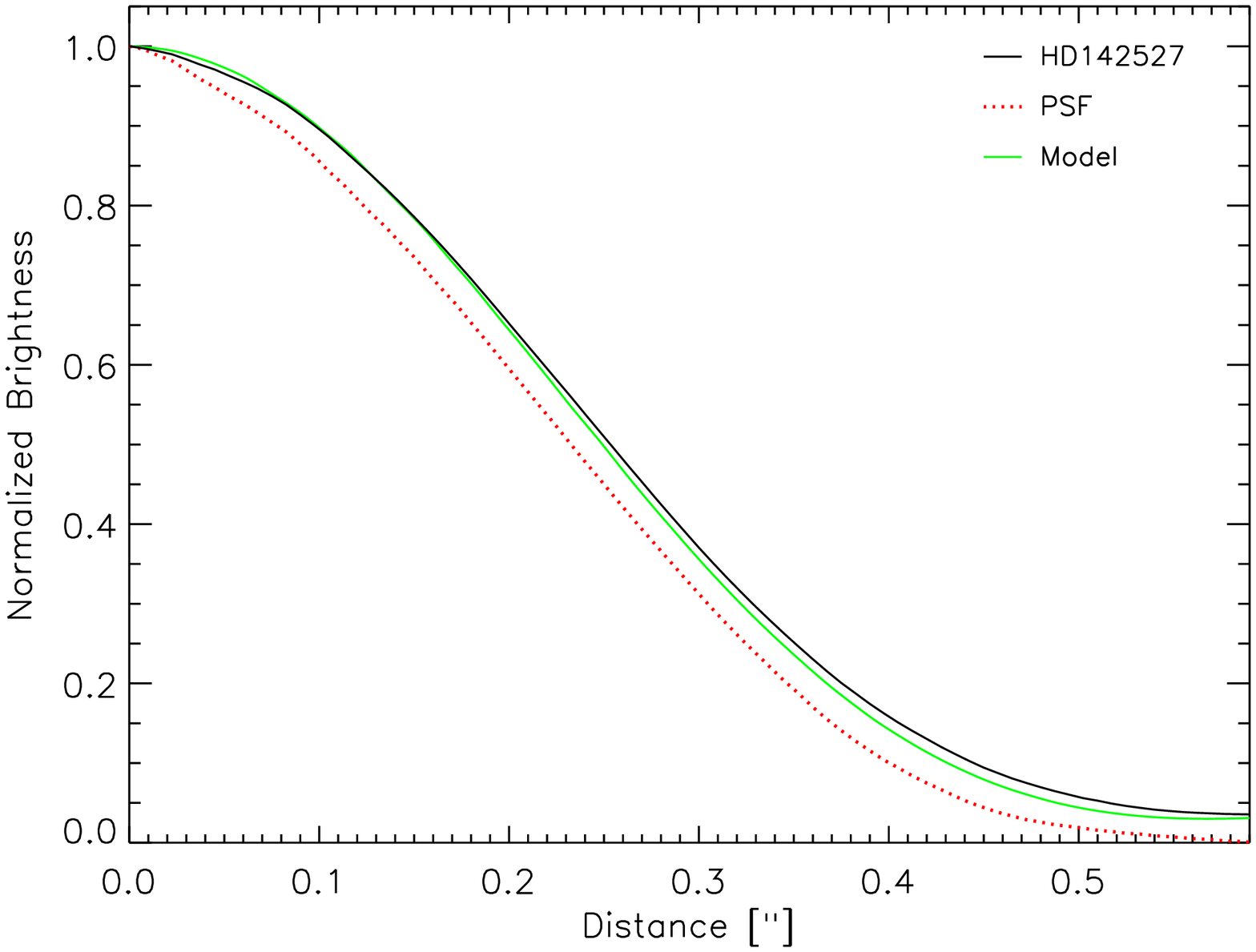}
\caption{Normalized mean radial surface brightness profile of the 
VISIR 18.72\,$\mu$m image (black) and the PSF (red). We also plotted
the result of our modeling effort (green; see
Sect.\,\ref{sec:model}). }
\label{fig:innerdisk}
\end{figure}
}

\newcommand{\decon}{
\begin{figure}[t!]
\includegraphics[width=\columnwidth]{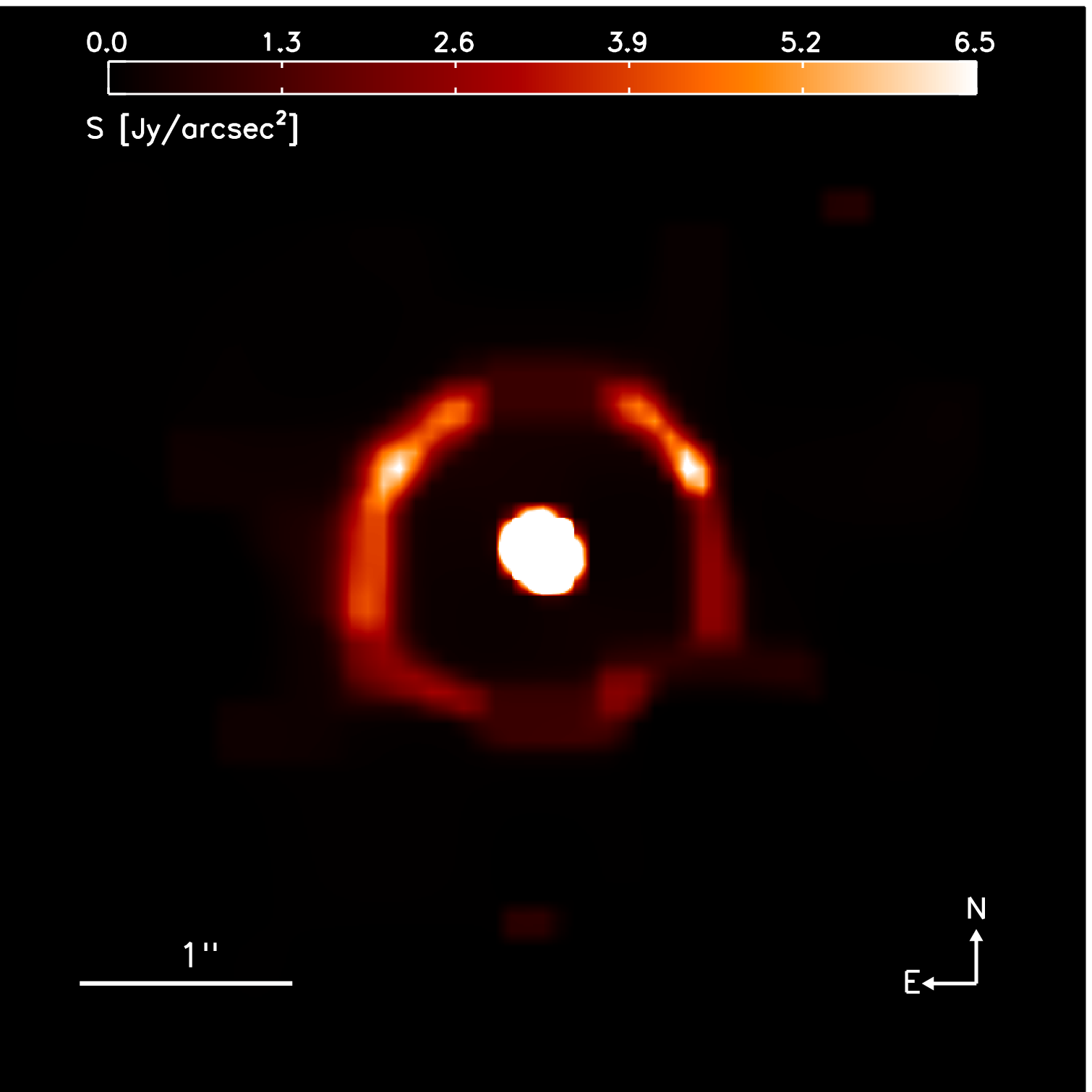}
\caption{Deconvolved VISIR Q-band (18.72~$\mu$m) image of 
HD\,142527. To zoom in on the lower surface brightness
we made a cut-off at 6.5\,Jy/arcsec$^2$. }
\label{fig:decon}
\end{figure}
}

\newcommand{\deconpro}{
\begin{figure}[t!]
\includegraphics[width=\columnwidth]{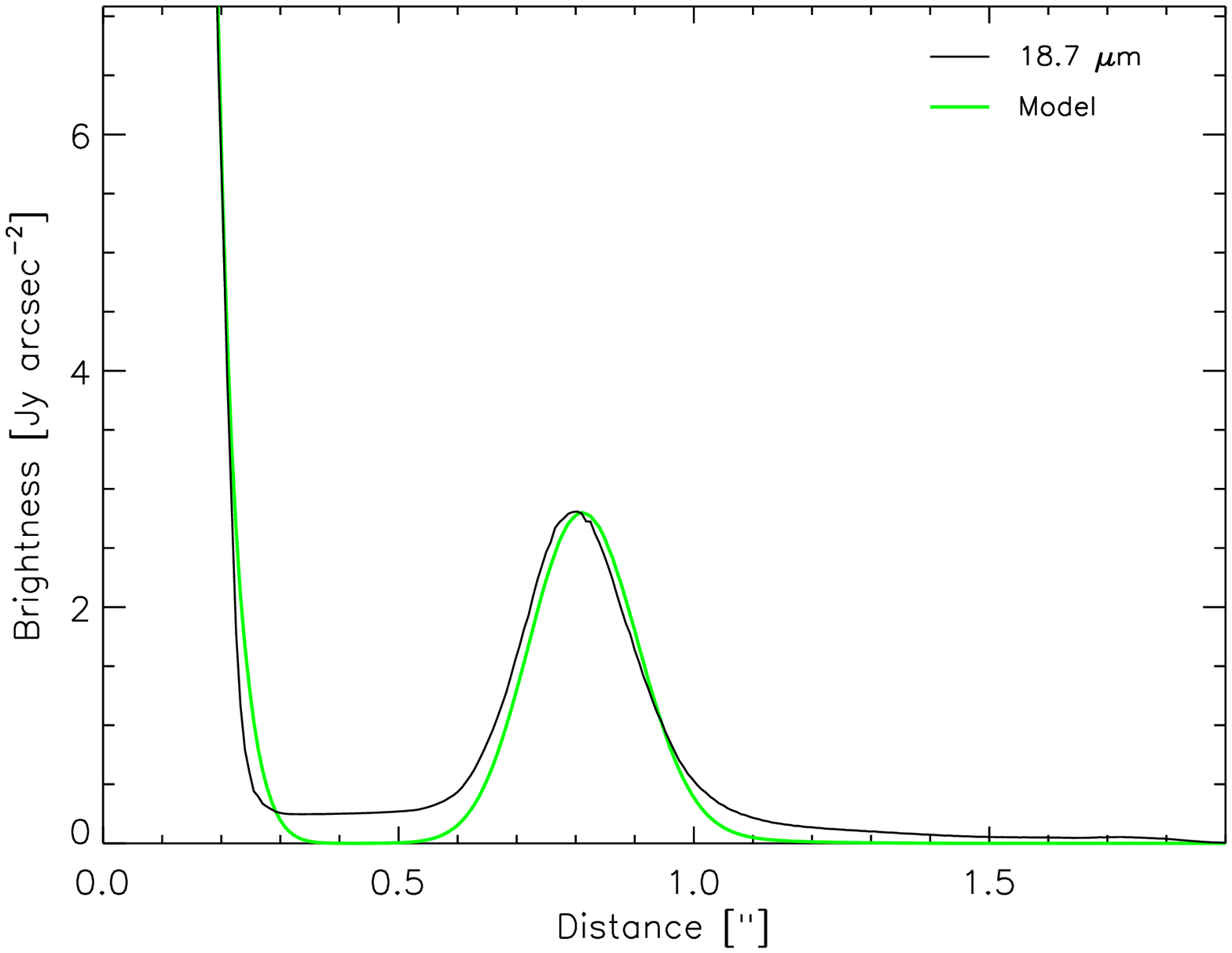}
\caption{Mean radial brightness profile as a function of distance 
to the star as retrieved from the deconvolved 18.72~$\mu$m image
(black). The result from the model image convolved with a Gaussian with
a FWHM of 0.1$\arcsec$ is overplotted in green.}
\label{fig:deconpro}
\end{figure}
}

\newcommand{\azprofile}{
\begin{figure}[t!]
\includegraphics[width=\columnwidth]{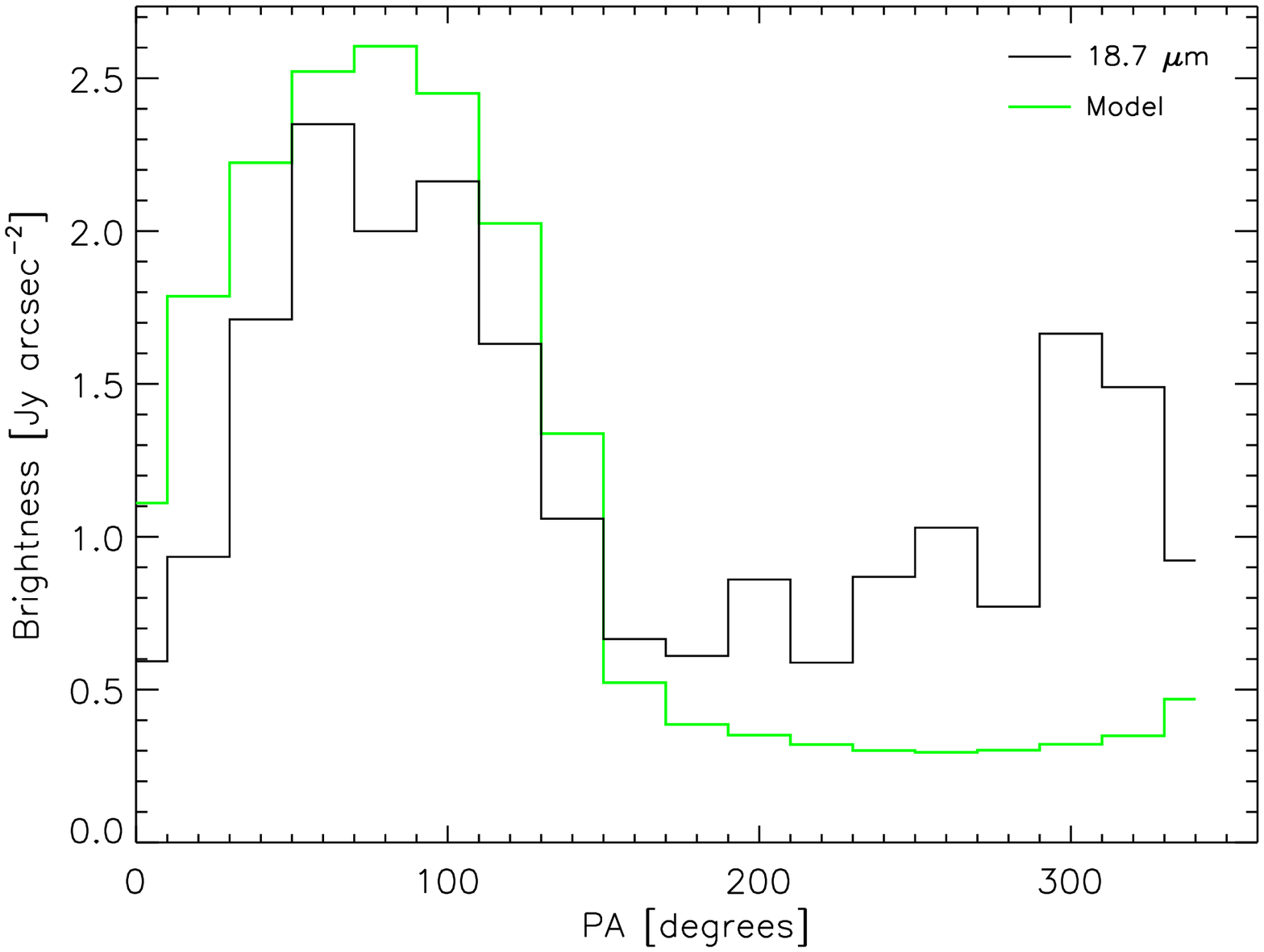}
\caption{Mean brightness profile of the outer disk (0.5$\arcsec < r 
< 1.1\arcsec$) as a function of the position angle (PA; east from
north). The result of the deconvolved 18.72\,$\mu$m image is plotted
in blue and the result of our model image is plotted in green.  }
\label{fig:az_profile}
\end{figure}
}

\newcommand{\VISIRSPC}{
\begin{figure}[t!]
\includegraphics[width=\columnwidth]{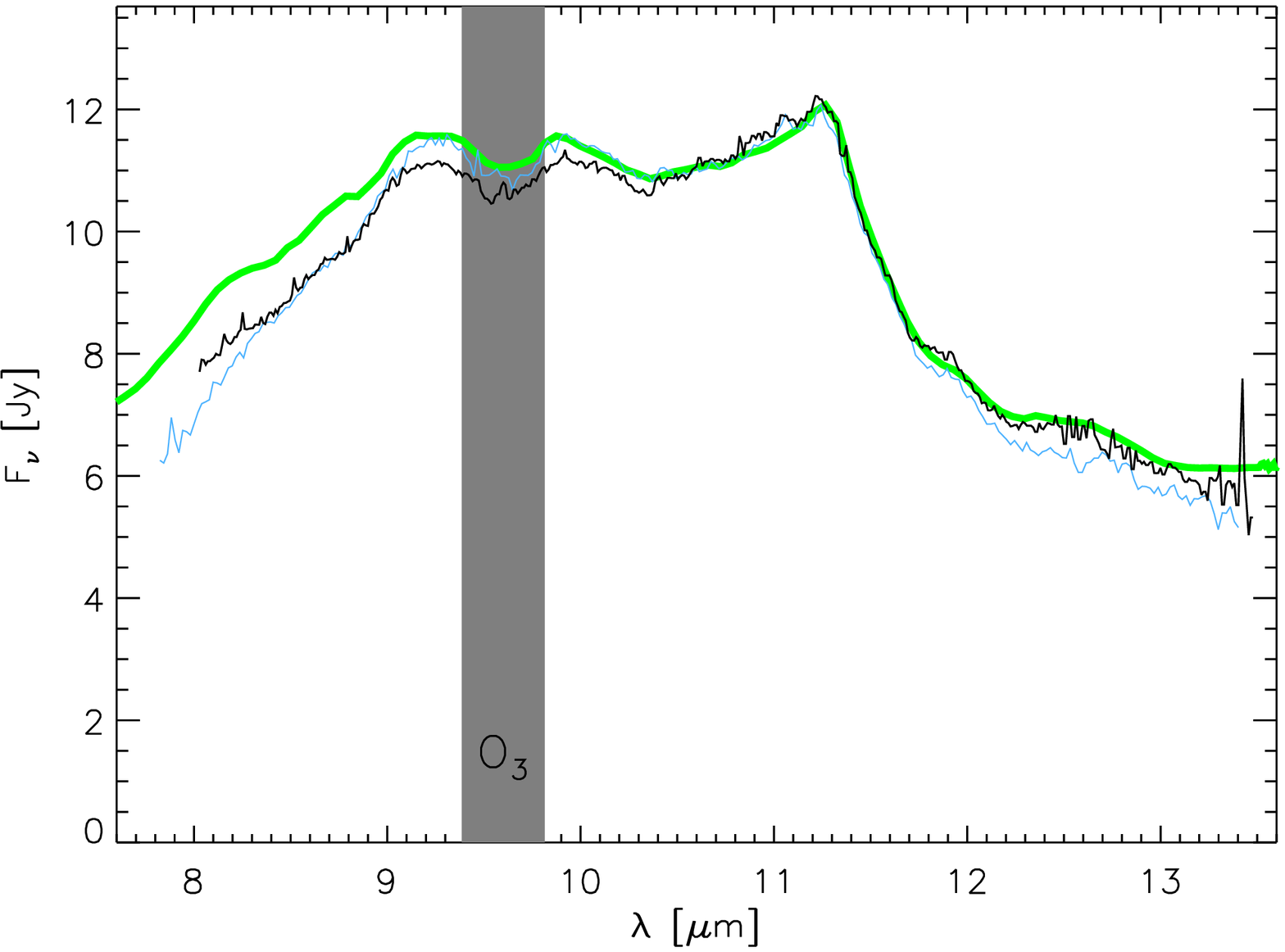}
\caption{VISIR N-band spectrum of HD\,142527 (black). Overplotted 
are the Spitzer spectrum (thick green) and the TIMMI2 spectrum (thin
blue).}
\label{fig:VISIRSPC}
\end{figure}
}

\newcommand{\QSPC}{
\begin{figure}[t!]
\includegraphics[width=\columnwidth]{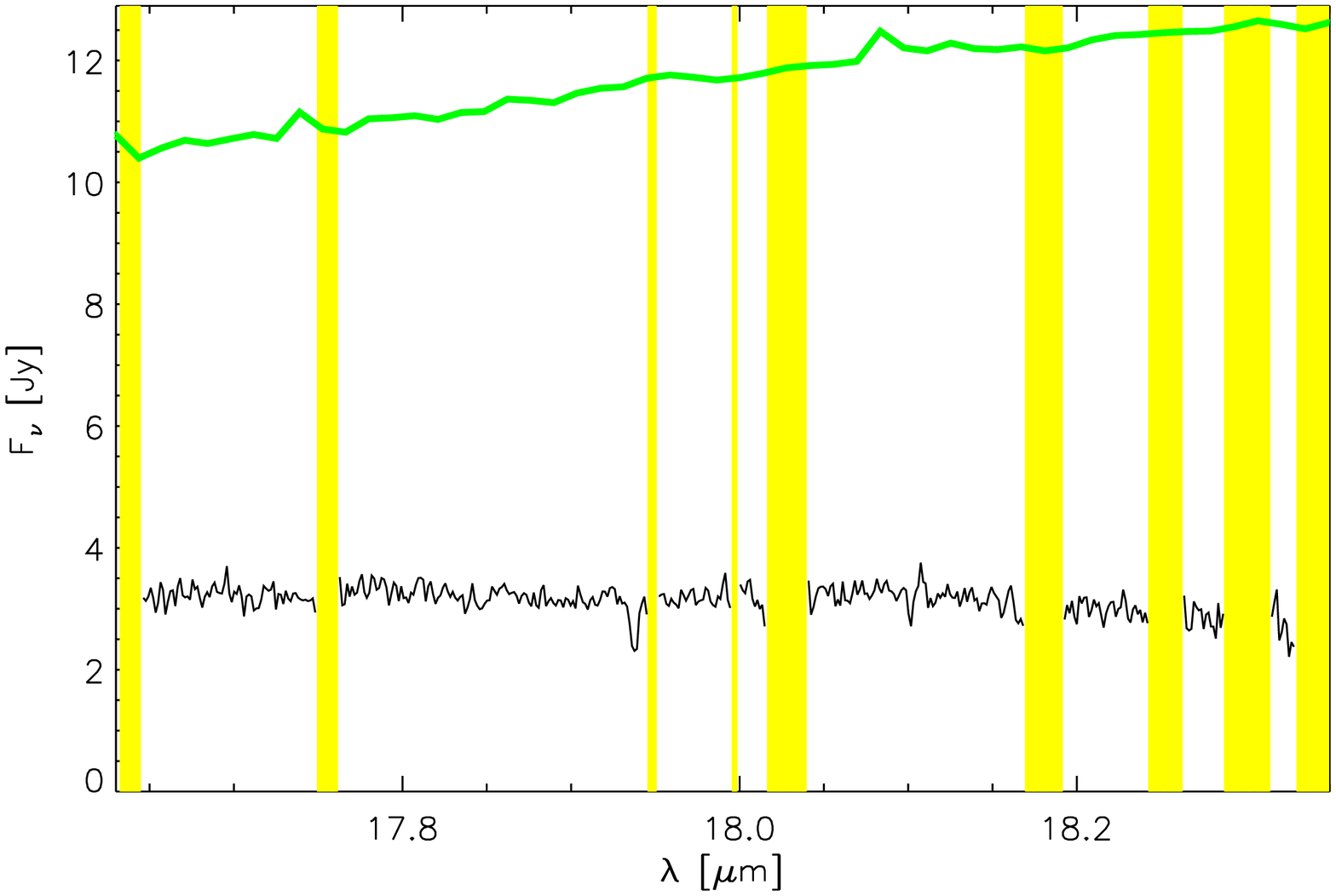}
\caption{VISIR Q-band spectrum of HD\,142527. Overplotted are 
the Spitzer spectrum (thick green) and the wavelengths of poor
atmospheric transmission (yellow bars).}
\label{fig:QSPC}
\end{figure}
}

\newcommand{\profile}{
\begin{figure}[t!]
\includegraphics[width=\columnwidth]{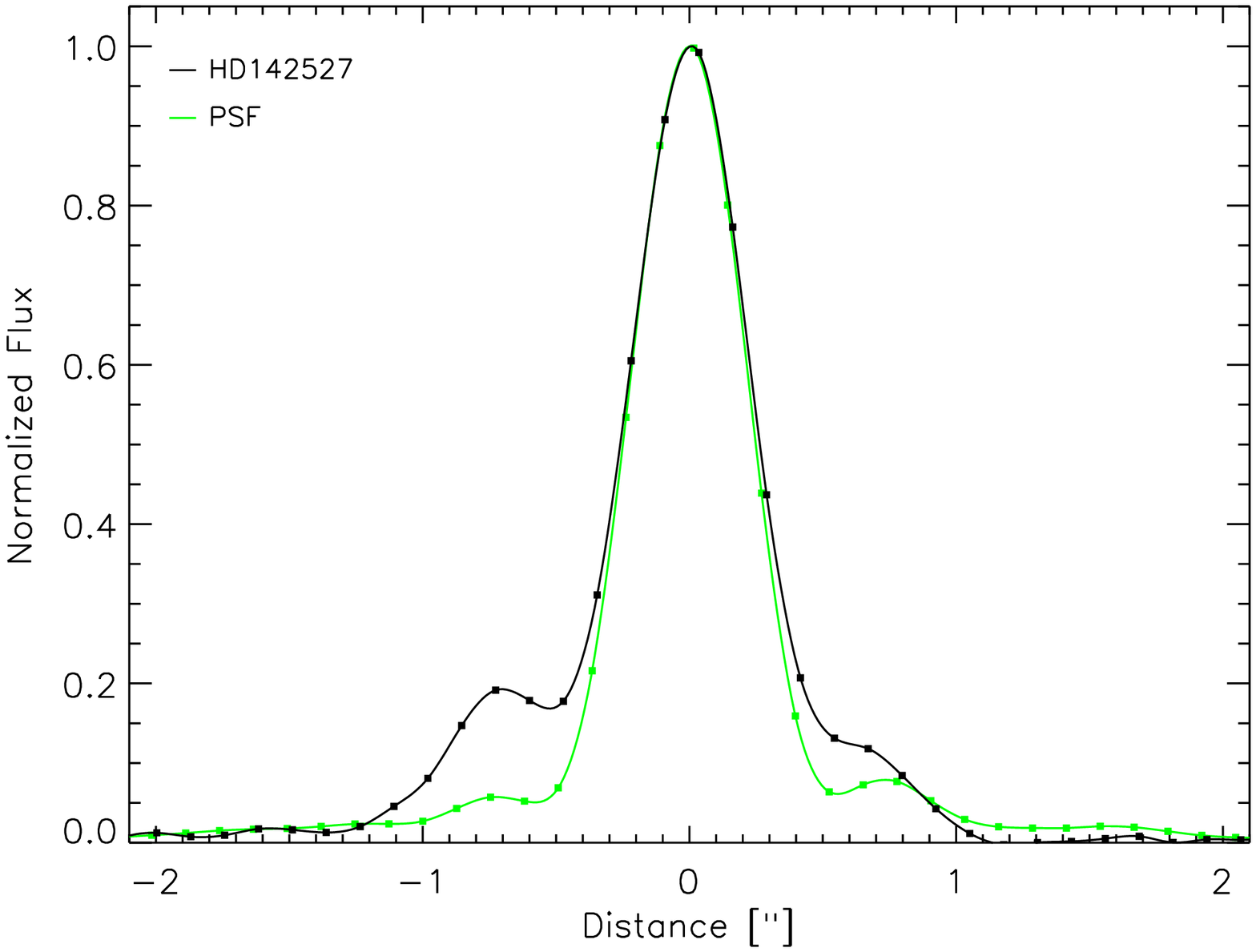}
\caption{Spatial emission profile of the Q-band spectrum at 
17.82\,$\mu$m ($\Delta \lambda$ = 0.15\,$\mu$m). Overplotted in 
green is the corresponding calibration measurement. }
\label{fig:profile}
\end{figure}
}

\newcommand{\MIDI}{
\begin{figure}[t!]
\includegraphics[width=\columnwidth]{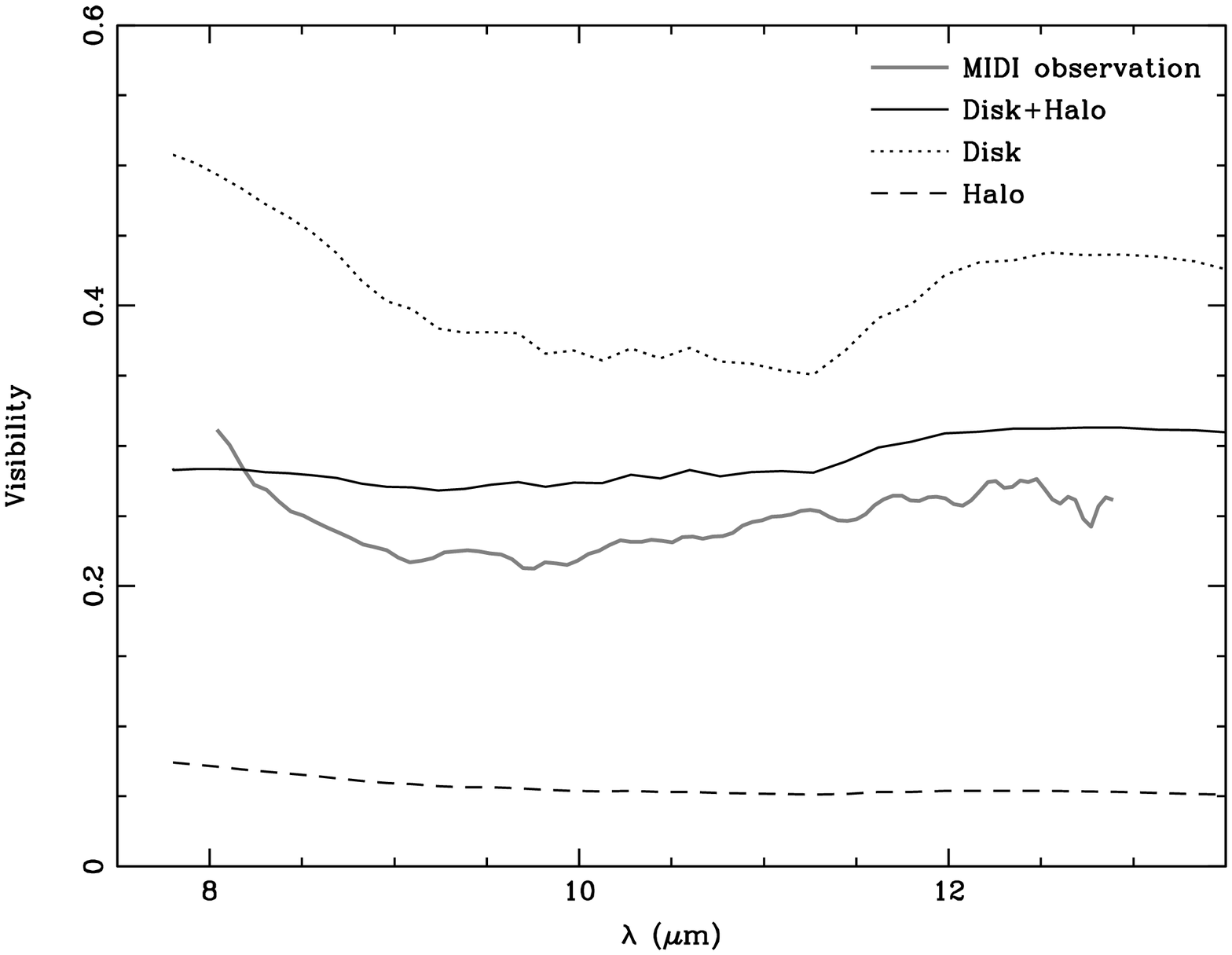}
\caption{MIDI visibilities compared with the different models of 
the inner disk region. The model that combines an optically thick disk
with an optically thin spherical halo fits the visibilities best. \ch{Typical
uncertainties on the MIDI visibilities are below 5\%.}
}
\label{fig:MIDI}
\end{figure}
}

\newcommand{\mspectra}{
\begin{figure}[t!]
\includegraphics[width=\columnwidth]{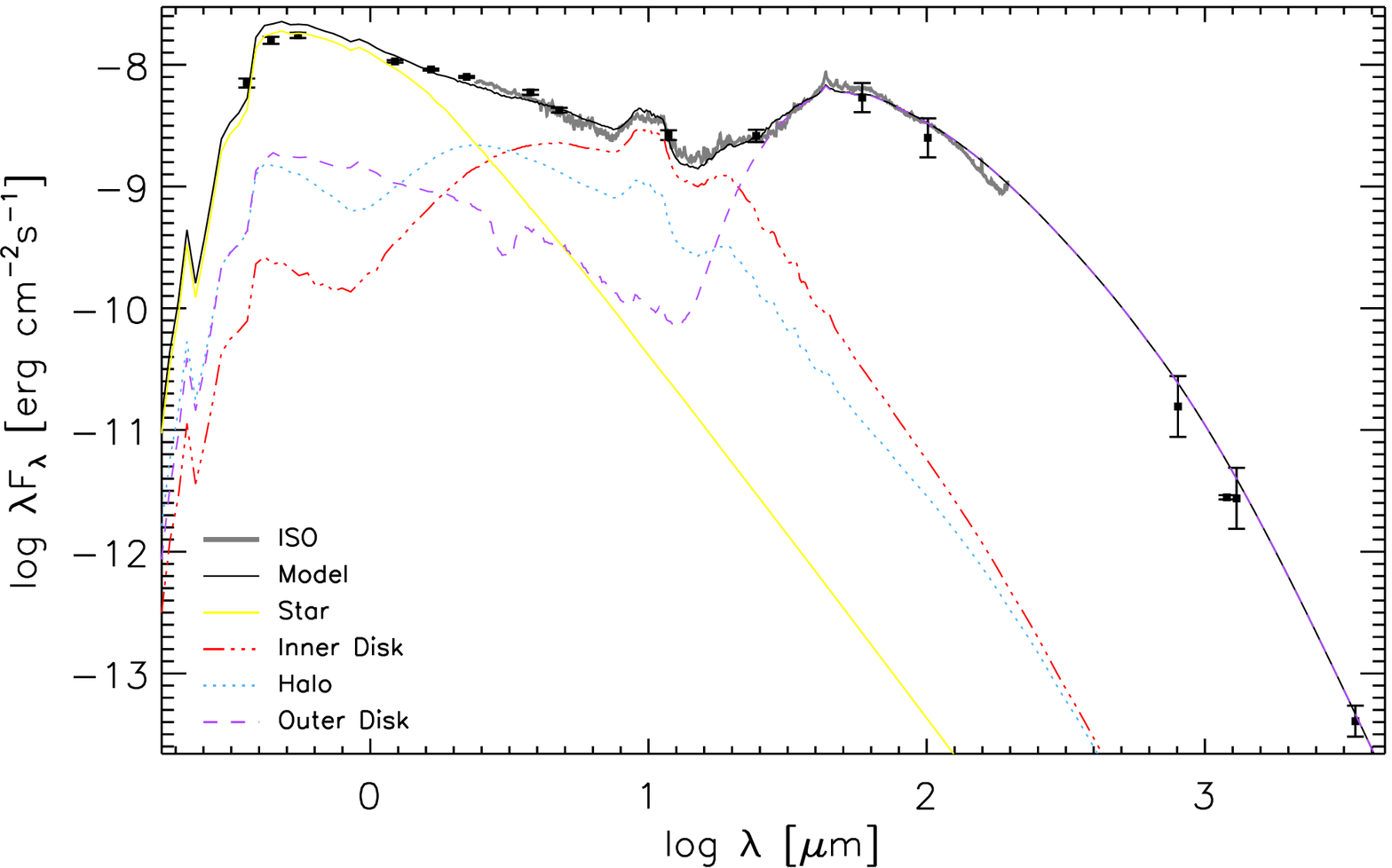}
\caption{Model fit (black) to the literature photometry (squares with 
error bars) and ISO spectrum (gray). \ch{The different flux components,
indicated in the bottom left, represent the photons whose last contact 
was with that particular model component.} }
\label{fig:mspectra}
\end{figure}
}

\newcommand{\modelima}{
\begin{figure}[t!]
\includegraphics[width=\columnwidth]{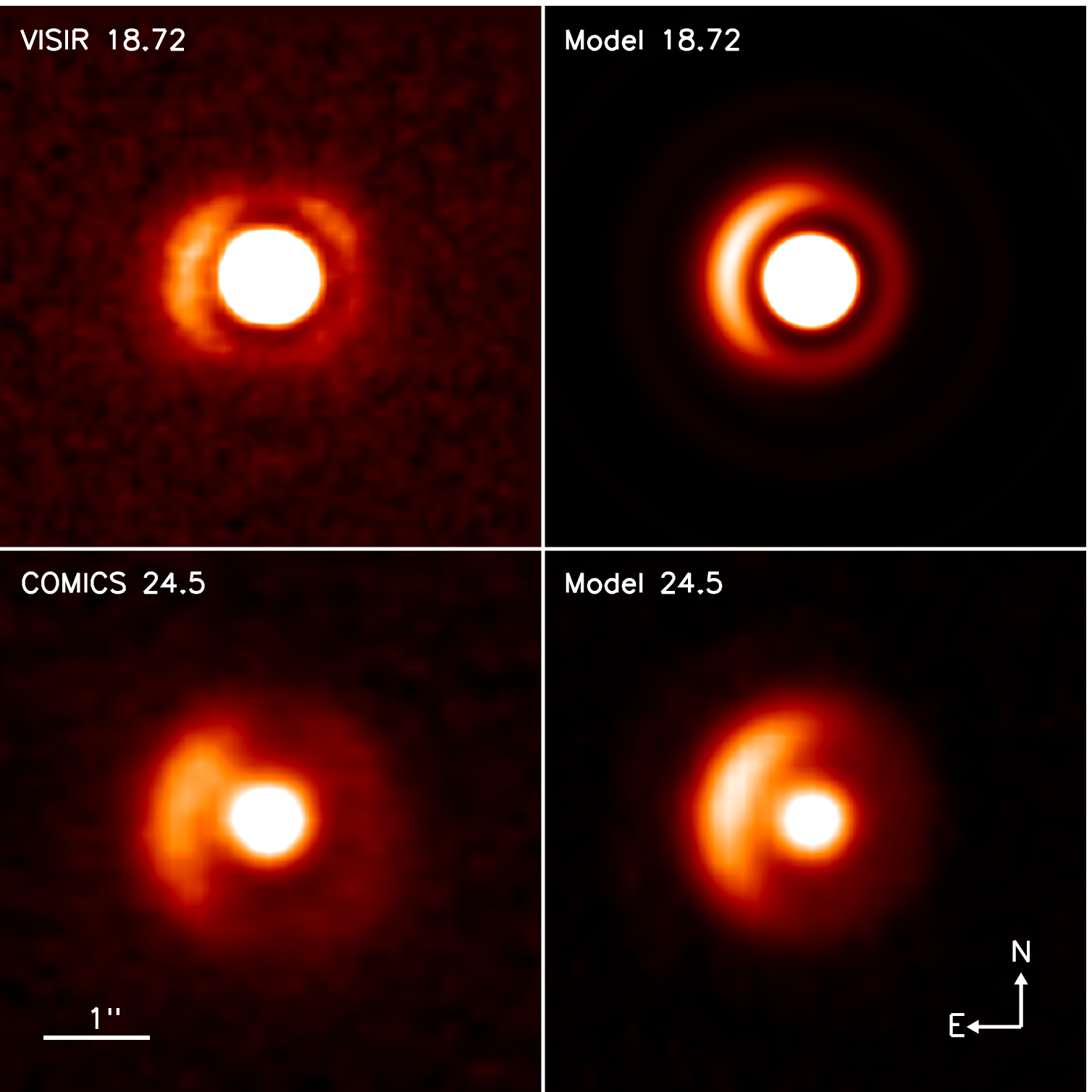}
\caption{Comparison of the VISIR 18.72\,$\mu$m (upper left), COMICS 
24.5\,$\mu$m (bottom left), and their modeled images (to the right). 
The model images were obtained by convolving the model output at
the specified wavelengths with the observed PSFs. For a clear 
comparison the observed and modeled images were displayed 
with the same lower and upper flux level cut-offs.}
\label{fig:modelima}
\end{figure}
}

\newcommand{\miprofile}{
\begin{figure}[t!]
\includegraphics[width=\columnwidth]{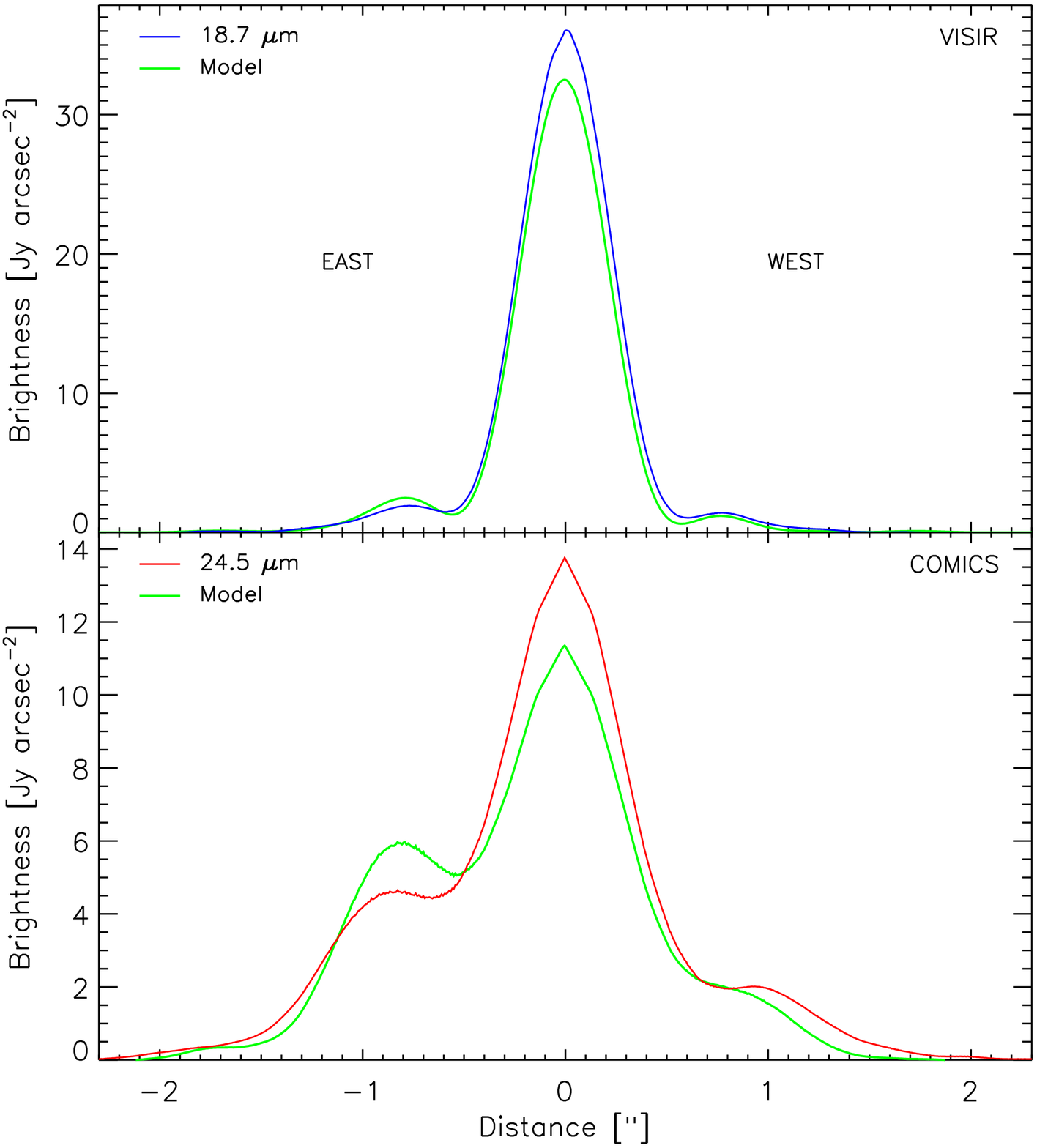}
\caption{Mean radial brightness profiles of the east 
($0^\circ<PA<180^\circ$) and west ($180^\circ<PA<360^\circ$) sides of
the VISIR 18.72\,$\mu$m image (blue) and the COMICS 24.5\,$\mu$m image
(red) compared with the profiles of the model images (green). }
\label{fig:miprofile}
\end{figure}
}

\newcommand{\picto}{
\begin{figure}[t!]
\includegraphics[width=\columnwidth]{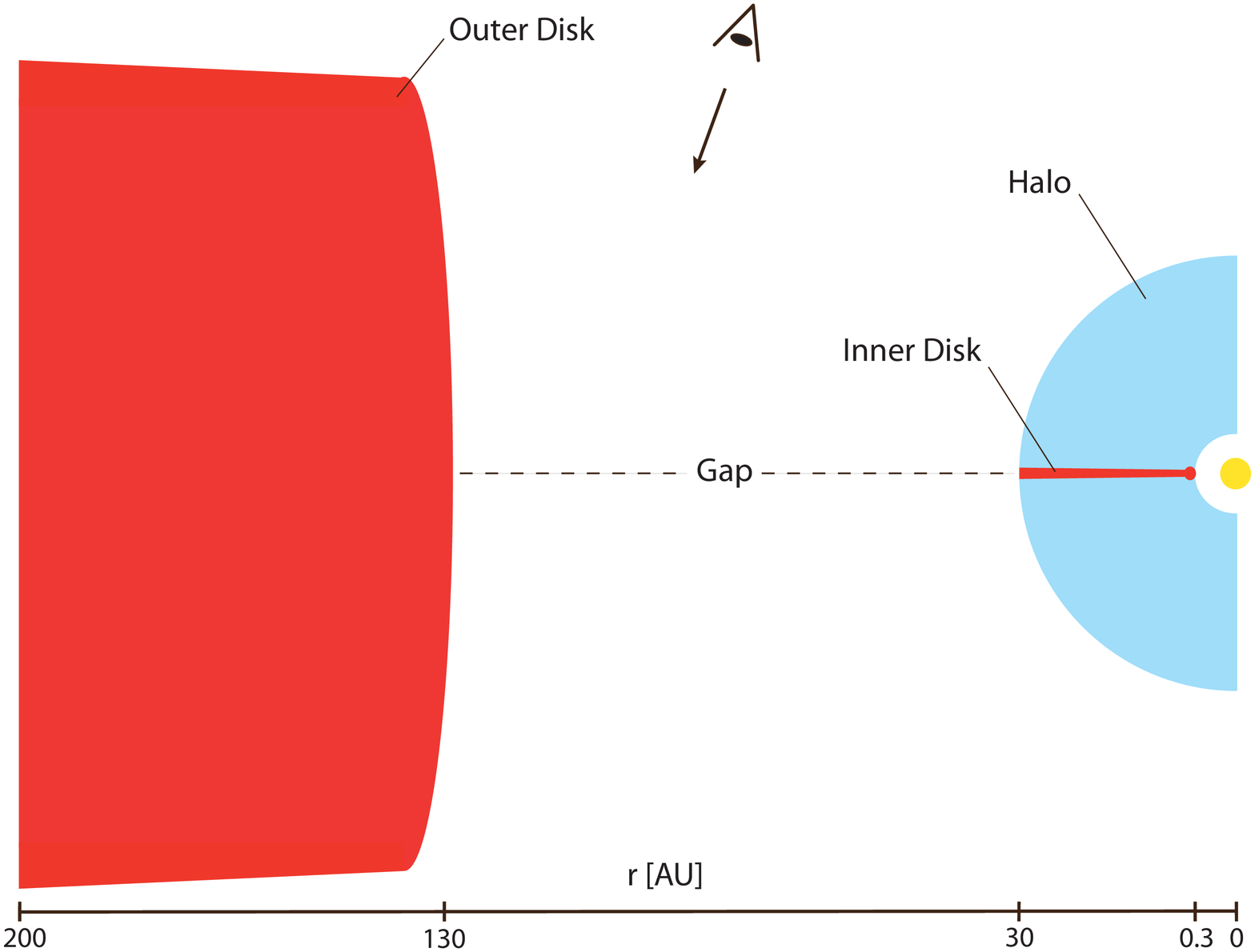}
\caption{Pictographic display of our model of the disk around HD\,142527.  }
\label{fig:picto}
\end{figure}
}

\newcommand{\ch}[1]{#1}
\newcommand{\ca}[1]{#1}
\newcommand{\ct}[1]{#1}

\begin{document}

\title{The complex circumstellar environment of HD\,142527. 
\thanks{Based on observations collected at the European Southern 
Observatory, Chile. Under program IDs: 075.C-0687A, 076.C-0266A and 
079.C-0286A.} 
}

\author{
A.P. Verhoeff \inst{1}
\and M. Min \inst{2}
\and E. Pantin \inst{3}
\and L.B.F.M. Waters \inst{4,1,5}
\and A.G.G.M. Tielens \inst{6}
\and M. Honda \inst{7}
\and H. Fujiwara \inst{8}
\and J. Bouwman \inst{9}
\and R. van Boekel \inst{9}
\and S.M. Dougherty \inst{10}
\and A. de Koter \inst{1,2}
\and C. Dominik \inst{1}
\and G.D. Mulders\inst{1,4}
}

\offprints{A.P. Verhoeff, \email{dr.verhoeff@gmail.com}}

\institute{Astronomical Institute ``Anton Pannekoek'', 
University of Amsterdam, P.O. Box 94249, 1090 GE Amsterdam, 
The Netherlands
\and Astronomical Institute, University of Utrecht, P.O. Box 80000, 
3508 TA Utrecht, The Netherlands
\and CEA/DSM/DAPNIA/Service d'Astrophysique, CE Saclay F-91191 Gif-sur-Yvette, France
\and SRON Netherlands Institute for Space Research, PO Box 800, 9700 AV Groningen, The Netherlands 
\and Institute for Astronomy, Catholic University Leuven,
Celestijnenlaan 200D, B-3001 Leuven, Belgium 
\and Leiden Observatory, Niels Bohrweg 2, 2333 CA Leiden, The Netherlands
\and Department of Information Science, Kanagawa University, 2946 Tsuchiya, Hiratsuka, Kanagawa 259-1293, Japan 
\and Department of Astronomy, School of Science, University of Tokyo, Bunkyo-ku, Tokyo 113-0033, Japan
\and Max Planck Institut f\"ur Astronomie, K\"onigstuhl 17, 69117 Heidelberg, Germany 
\and Hertzberg Institute for Astrophysics, DRAO, P.O. Box 248, Penticton, Canada
}

\date{Received 7 May 2010 / Accepted 14 January 2011}
 
\abstract {The recent findings of gas giant planets around young
A-type stars suggest that disks surrounding Herbig Ae/Be stars
will develop planetary systems. An interesting case is HD\,142527, for
which previous observations revealed a complex circumstellar
environment and an unusually high ratio of infrared to stellar
luminosity. Its properties differ considerably from other Herbig
Ae/Be stars. This suggests that the disk surrounding HD\,142527 is in
\ch{an uncommon evolutionary stage.}}
{We aim for a better understanding of the geometry and
evolutionary status of the circumstellar material around the Herbig
Ae/Be star HD\,142527. }
{We map the composition and spatial distribution of the dust around
HD\,142527. We analyze SEST and ATCA millimeter data, VISIR N and
Q-band imaging and spectroscopy. We gather additional relevant data
from the literature. We use the radiative transfer code MCMax to
construct a model of the geometry and density structure of the
circumstellar matter, which fits all of the observables
satisfactorily. }
{We find that the disk of HD\,142527 has three geometrically distinct
components \ct{separated by a disk gap running from 30 to 130\,AU. There
is a geometrically flat inner disk running from 0.3\,AU up to 30\,AU; an 
optically thin \ch{halo-like component} of dust 
in the inner disk regions; and a massive self-shadowed outer disk running 
from 130\,AU up to 200\,AU. We derived a total dust mass in small grains 
of 1.0$\cdot$10$^{-3}$\,M$_{\odot}$ and a vertical height of the inner wall
of the outer disk of h = 60\,AU. Owing to the gray extinction of the 'halo' we 
obtained new stellar parameters, including a stellar luminosity of 
20$\pm$2\,L$_{\odot}$ and age of 10$^{6.7\pm0.4}$\,yr.}}
{\ct{We find that the disk surrounding HD\,142527 is highly evolved despite the 
relatively young age of the star. The peculiar disk geometry 
can explain the extreme IR reprocessing efficiency of the disk. Furthermore, the 
geometry, the large disk mass, and the highly processed dust composition are 
indicative of on-going planet formation.} }


\maketitle

\section{Introduction}
Herbig Ae/Be (HAeBe) stars are intermediate mass, pre-main sequence
(PMS) stars with an infrared (IR) excess and emission lines. The
observed IR excess is caused by circumstellar dust confined to a
disk. Based on the amount of far-IR excess emission the geometry of
these disks is considered to be either flaring or flattened,
classified by \cite{2001A&A...365..476M} as group I or II
respectively. The composition, structure and evolution of these disks
has been extensively studied because they are generally believed to be the
sites of ongoing planet-formation. Recent discoveries of gas giant
planets in wide orbits around intermediate mass young stars
\citep{2008Sci...322.1348M, 2008Sci...322.1345K,2009A&A...493L..21L}
provide exciting support for this interpretation of disks around
Herbig Ae/Be stars.  In this study, we focus on the F6IIIe star
HD\,142527. This star is particularly interesting because its
\ch{extremely large} IR and millimeter excess is hard to understand in
terms of a passive disk model (see \citealt{2003A&A...398..607D}).

The disk of HD\,142527 has been imaged in scattered light
\ch{\citep{2006ApJ...636L.153F}} and in the thermal IR
\citep{2006ApJ...644L.133F}, showing that the disk extends to several
\ct{hundred} AU with a prominent disk gap at a distance of about 100\,AU.  The
dust in the disk is highly processed. Both the scattered light and
thermal IR images show evidence for grain growth: the inferred typical grain sizes
are 1-2\,$\mu$m. This picture is also seen in the thermal IR
spectrum of HD\,142527. Van Boekel et al. (2004a) show that the
silicates in the inner disk, spatially resolved with the MIDI
instrument at the Very Large Telescope Interferometer, are fully
crystalline. Even the integrated disk spectrum of HD\,142527 has a
crystallinity of more than 20\%. The outer cold dust is characterized
by crystalline water ice and possibly hydro-silicates
\citep{1999A&A...345..181M}, again pointing to a highly processed dust
environment. Recently, \cite{2009ApJ...690L.110H} showed evidence for
crystalline water ice in the near-IR scattered light spectrum of the
disk. \cite{2008Ap&SS.313..101O} showed that the cold dust is
distributed in an arc-like structure, which adds to the complexity of the
source. 

Clearly, the disk of HD\,142527 is in an interesting evolutionary
phase. Its unusual properties might be caused by giant planet
formation in the outer disk. In particular we are interested in
understanding the very large far-IR excess, the geometry of the disk
and the mineralogy of the dust. For this purpose we gathered a
comprehensive data set on this object, consisting of SEST and ATCA
millimeter photometry, a Spitzer-IRS spectrum, ISO-SWS and LWS
spectra, VISIR N and Q-band imaging, VISIR N and Q-band spectroscopy,
and optical, infrared and millimeter photometry.  In this paper, we
focus on the SED and the disk geometry. In a future paper (Min et al.,
in preparation) we will study the dust mineralogy. This paper is
structured as follows: \ct{In Sect.\,\ref{sec:parameters} we motivate the 
adopted stellar parameters.} \ch{Section}\,\ref{sec:obs} describes the
observations and the data reduction. In Sects.\,\ref{sec:anamilli}
through \ref{sec:vspcana} we carry out an in-depth analysis of these
data-sets. In Sect.\,\ref{sec:model} we construct a comprehensive
model of the disk geometry using a Monte Carlo radiative transfer
code \ch{and address the discrepancies between model and 
observations. The discussion in Sect.\,\ref{sec:disc} considers} the
emerging geometrical picture of the circumstellar matter and possible
explanations for the peculiarities. \ct{Section\,\ref{sec:conclusions} 
summarizes our main conclusions.}

\section{Stellar parameters}
\label{sec:parameters}
\parameters 
HD\,142527 was cataloged as an F6III star by
\cite{1978mcts.book.....H}. \cite{1976ApJS...30..491H} noted its
emission line nature and \cite{1996A&A...315L.245W} classified it as a
Herbig star. With Hipparcos measurements, \cite{1998A&A...330..145V}
placed it at a distance of 200$^{+60}_{-40}$\,pc, but we consider
the association by \cite{2004A&A...426..151A} to the star-formation
region Sco~OB2-2 to be more reliable and therefore adopt the distance of
145\,pc (\citealt{1999AJ....117..354D}), \ch{which is within 2 sigma
  of the Hipparcos distance}. By comparing a
\cite{1991ppag.proc...27K} model for the photosphere associated with
the given spectral type to the literature photometry, we obtained an
optical extinction of $A_{\rm V}$ = 0.60$\pm$0.05 and a stellar
luminosity of $L$ = 15$\pm$2\,L$_\odot$. However, our modeling of the dust 
(see \ch{Sect.\,\ref{sec:fitting}})
demonstrates the presence of a gray extinction component. Taking this
component into account, we derive a stellar luminosity of $L$ =
20$\pm$2\,L$_\odot$. A stellar radius of $R_*$ =
3.8$\pm$0.3\,R$_\odot$ was derived from the effective temperature and
the luminosity. The position in the Hertzsprung-Russell diagram was compared with the PMS
evolution tracks of \cite{2000A&A...358..593S}, which resulted in a
stellar mass of $M$ = 2.2$\pm$0.3\,M$_\odot$ \ch{and an age of 
10$^{6.7\pm0.4}$\,yr}. Table\,\ref{tab:parameters} summarizes the stellar 
parameters.

\section{Observations and data reduction}
\label{sec:obs}
\subsection{SEST}
HD\,142527 was observed with the 15\,m Swedish/ESO sub-mm Telescope
(SEST) in August 2003. The 37-channel SEST imaging bolometer array
\citep[SIMBA,][]{2001Msngr.106...40N} was used to observe the
continuum dust emission at 1.2\,mm. The observations were performed in
fast mapping mode, and maps were made of 400$\arcsec$ by
500$\arcsec$ in azimuth and elevation respectively. The atmospheric
transparency was monitored by performing regular sky-dip measurements,
the absolute flux calibration was established by observing Uranus. In
total, six maps of HD\,142527 were made, and the source was clearly
detected in all of them. We determine the \ct{flux} of
HD\,142527 at 1.2\,mm to be 1.12$\pm$0.02\,Jy.

\ATCA
\mmobs
\subsection{ATCA}
Observations at 3.476\,mm were obtained with the Australia Telescope
Compact Array (ATCA) on the night of June 11, 2002 in the \ch{east-west 352\,m}
configuration using a bandwidth of 2.3\,mm \ch{(see
  table\,\ref{tab:mmobs})}.  Only three antennae out of the six in the
full array were available (numbers 2, 3, and 4). The resolution was
16.2$\arcsec$ by 2.9$\arcsec$. Observations on the target source were
interleaved with observations of the bright, nearby \element{SiO}
maser source, IRSV\,1540, as a phase-reference calibrator, with a
cycle of two minutes on the phase-reference followed by five minutes
on the target. Every half hour $T_{\rm sys}$ observations were made
and every hour a new pointing solution derived using the
phase-reference. \ch{The data reduction was completed using the {\sc
MIRIAD} data reduction package (\citealt{1995ASPC...77..433S}).}
The absolute flux scale and bandpass were calibrated using an
observation of Uranus. The flux of Uranus was determined to be
7.0\,Jy. From the variance in flux over time of the sources IRSV\,1540
and 1057-797, the flux scale appears to be constant to within
10\%. \ct{The measured central position matches the optical position of 
HD\,142525 to a sub-arcsecond precision.} Fitting a Gaussian to the source 
shown in Fig.\,\ref{fig:ATCA}
suggests that the source is marginally resolved, \ct{but the visibility 
data are too noisy to obtain a size estimate.} The derived peak flux 
is 43.1$\pm$5.4\,mJy with a spatially integrated flux of
47.1$\pm$6.5\,mJy. This last number was previously quoted by
\cite{2004A&A...422..621A}\footnote{In that paper the wavelength of
the ATCA data-point was incorrectly stated as 2.9\,mm.} as
unpublished.

\logima
\subsection{VISIR imaging}
\label{sec:VIMA}
During the nights of April 28 and June 30, 2005 N and Q-band imaging
of HD\,142527 was performed with VISIR, the VLT Imager and
Spectrometer for mid-IR (\citealt{2004Msngr.117...12L}). The
observations were obtained in visitor mode under the VISIR GTO program
on circumstellar disks. The chosen filters were the SiC and Q2, which
have central wavelengths at 11.85 and 18.72\,$\mu$m and half-band
widths of 2.34 and 0.88\,$\mu$m respectively. Standard chopping and
nodding was applied (at 0.25 and 0.033\,Hz) to suppress the dominating
atmospheric background emission. All the imaging data were obtained
using a pixel scale of 0.075$\arcsec$. The observing conditions during
both nights were good: the airmass was below 1.1 and the optical
seeing was around 0.8$\arcsec$. For the imaging in the SiC filter the
standard stars HD\,139127 (12.75\,Jy) and HD\,186791 (57\,Jy) were
bracketing the science observation and were used to estimate the
photometry and the associated errors. The achieved sensitivity was
estimated to be 10\,mJy/10\,$\sigma$1h averaged over the two reference
stars. For the imaging in the Q2 filter the reference star, HD\,139127
(4.95\,Jy), was used for PSF and photometric purposes.  It was
selected from the database of VISIR standard stars based on the flux
level and distance to the science target on the sky. The achieved
sensitivity was estimated to be 70\,mJy/10\,$\sigma$1h in the Q2
filter.  Table\,\ref{tab:logima} summarizes the observational
details. For the reduction of the data we used a dedicated
pipeline, which corrects for various instrumental signatures (see
\citealt{2008SPIE.7014E..70P,2009svlt.conf..261P, Pantin2010}).

\logmr
\subsection{VISIR spectroscopy}
Long-slit spectra were taken with VISIR in the N-band and the
Q-band. For both observations standard parallel chopping and nodding
with a chopper throw of 8$\arcsec$ was applied to correct for the
atmospheric background. \ch{In table\,\ref{tab:logmr} we display the 
observational log.}

The N-band spectrum was observed in the low-resolution mode
\ch{(R$\sim$300)} on the night of April 27, 2005. The 0.4$\arcsec$
slit was employed in a standard north-south orientation. The entire
10\,$\mu$m feature was covered in four separate settings. The
observing conditions were favorable; the average airmass was $\sim$1.1
and the optical seeing was 0.7$\arcsec$.  \ch{The standard star
HD\,139127 was observed immediately before or after the science
measurements} for the correction of the atmospheric absorption. The
N-band spectrum of HD\,142527 was observed as part of a VISIR GTO
program on circumstellar disks, in which the 10\,$\mu$m feature of a
sample of 17 Herbig Ae stars was studied \citep{2010A&A...HAEBE}.

The Q-band spectrum was observed in the medium-resolution mode
\ch{(R$\sim$3000)} on the night of June 30, 2007. The 0.4$\arcsec$
slit was employed in an inclined orientation with $PA$ =
67$^\circ$. The 18\,$\mu$m forsterite feature was mapped in two runs
centered on 17.82\,$\mu$m and 18.18\,$\mu$m.  The observing conditions
were fair; the average airmass was $\sim$1.1 and the optical seeing
was 1.4$\arcsec$. The standard star HD\,148478 was observed directly
after the science measurement for the correction of the atmospheric
absorption.  

The first steps of the data reduction are identical to the VISIR image
reduction (see Sect.\,\ref{sec:VIMA}). The resulting reduced spectral
images contain the two-dimensional spectra in counts/s. These images
were corrected for slit curvature and optical distortions. The spectra
were extracted using the optimal extraction technique of
\cite{1986PASP...98..609H}. The wavelength calibration was performed
by correlating an ALTRAN/HITRAN model with the simultaneously measured
sky spectra. See \cite{Pantin2010} for a more detailed description of
the reduction process.

The N-band and Q-band spectrum were corrected for atmospheric
extinction using one standard star and a simple correction for the
airmass difference. The formalism is defined as
\begin{equation} 
I_{\rm \lambda,s} = S_{\rm \lambda,s} \left( \frac{I_{\rm \lambda,c}}{S_{\rm \lambda,c}} \right)^\frac{m_{\rm s}}{m_{\rm c}},
\end{equation}
where $S_{\rm \lambda,c}$ and $S_{\rm \lambda,s}$ are the
observed counts of the calibration and the science measurements,
$m_{\rm c}$ and $m_{\rm s}$ are the airmass of the measurements, and
$I_{\rm \lambda,c}$ and $I_{\rm \lambda,s}$ are the
\ch{\cite{1999AJ....117.1864C} model} of the calibrator and the
resulting spectrum.

\phot
\subsection{Additional data}
A low-resolution Spitzer-IRS spectrum of HD\,142527 was taken from
\ch{\cite{2010ApJ...721..431J}}. The ISO spectrum was taken from
\cite{2003A&A...398..607D}. N-band interferometric visibilities
observed with MIDI were taken from \cite{2004Natur.432..479V}.
Photometric data points were taken from the literature \ch{and are listed in
table~\ref{tab:phot}. The optical photometry of
\cite{1998A&A...331..211M} were converted to the standard Johnson
UBV system using the transformations of \cite{1988A&A...206..357R}.}
The Subaru/Comics 24.5\,$\mu$m image was taken from
\cite{2006ApJ...644L.133F} and the TIMMI2 N-band spectrum from
\cite{2004A&A...418..177V}.

\section{Analysis of the millimeter data}
\label{sec:anamilli}
The two measurements at 1.2 and 3.5\,mm together with two more
measurements at 0.8 and 1.3\,mm from the literature 
(see table~\ref{tab:phot}) give
us \ct{an opacity index $\beta$ ($\kappa \propto \lambda^{-\beta}$)}. Assuming the four data-points are in
the Rayleigh-Jeans limit of the SED \ct{($\beta = \alpha - 2$)}, we obtain $\beta=1.0\pm0.1$.
Compared with other HAeBe stars this value is higher than average.
\cite{2007prpl.conf..767N} find that computed over the wavelength
interval 1.3-7\,mm, the opacity index $\beta$ is $<1$ for about 60\%
of the objects.  \cite{2004A&A...422..621A} determined the spectral
index in observations from 0.35 to 2.7\,mm and found an average of
$<\beta>$ = 0.6 for 13 group I sources. At mm wavelengths the SED is
thus steeper than that of most HAeBe stars. Considering that $\beta$ =
1.7 for the ISM (\citealt{2001ApJ...548..296W}) and assuming an optically 
thin disk at millimeter wavelengths, we can conclude that the
dust around HD\,142527 has evolved, but that the average grain-size is
relatively small compared with most HAeBe stars.

From the 1.2\,mm observation, we can derive a dust mass using the
formalism of \cite{1983QJRAS..24..267H} for an optically thin cloud
\begin{equation}
M_{\rm dust} = \frac{F_{\nu}\,d^2}{B(\nu,T)}\frac{4\,a\,\rho}{3 Q_{\nu}}, 
\end{equation}
with $a$ the grain radius and $\rho$ the grain density, $Q_{\nu}$ the
emissivity, and $d$ the distance to the star. Considering the
emissivity per unit dust mass $\kappa_{\rm \nu}$ =
$3Q_\nu/4\,a\,\rho$, taking $\kappa_{\rm \nu}$\,(250\,$\mu$m) =
10\,cm$^2$\,g$^{-1}$ from \cite{1983QJRAS..24..267H} and extrapolating
with $\beta$ = 1.0 we obtain $\kappa_{\rm \nu}$\,(1.2\,mm) =
2\,cm$^2$\,g$^{-1}$, which is commonly used in the literature and
should be correct within a factor of 4 (see
e.g. \citealt{1990AJ.....99..924B}). We assume an average dust
temperature of $T_{\rm dust}$ = 40\,K, which is not a large source of
uncertainty, because in the Rayleigh-Jeans limit \ch{the derived mass
  scales inversely proportional with the temperature.} Using the
distance of $d$ = 145\,pc, we calculate a total dust mass \ct{in small grains}
of $M_{\rm dust}$ = $1\cdot10^{-3}$\,M$_{\odot}$. This value is consistent with
the model we present in Sect.\,\ref{sec:model}.

\slitpos
\innerdisk
\section{VISIR imaging analysis}
The photometric analysis yields integrated fluxes for HD\,142527 of
8.8$\pm$0.6\,Jy and 14.6$\pm$1.5\,Jy in the SiC (11.85\,$\mu$m,
spectral resolution R~=~5) and Q2 (18.72\,$\mu$m, R~=~20) filters
respectively. Because the emission of the star is negligible at these
wavelengths, these fluxes can be attributed to the circumstellar
disk. We searched for extended emission components at both wavelengths
with a central component subtraction technique at a sub-pixel (1/10)
precision level. The PSF in both filters was derived from the
observation of standard stars just before or after HD\,142527.  In the
SiC filter, no visible extension is detected. This constrains the
angular extent of the central component to $<$ 0.35$\arcsec$ in FWHM
(50\,AU at 145\,pc). In the Q-band, extended emission is clearly
detected. A visual inspection of the observed image in
Fig.\,\ref{fig:slitpos} demonstrates that the emission is divided
between a compact central component and an extended region. We find
that the emission in the central component is also marginally
spatially resolved (see Fig.\,\ref{fig:innerdisk}). A quadratic
subtraction of the PSF leads to a FWHM of the spatial emission profile
of 0.21$\arcsec\pm$0.04, which corresponds to 30\,AU at a distance of
145\,pc.

\contour The extended component that is visible after central component
subtraction (figure \,\ref{fig:contour}) has an elliptical shape. An
ellipse fitting of the outer boundary isophotes indicates a position
angle of 110$^\circ\pm$20$^\circ$ and an aspect ratio of 1.3 (i =
40$^\circ\pm$20$^\circ$ from face-on). The integrated flux in the
residuals after central component subtraction is 2.9$\pm$0.5\,Jy and
the integrated flux of the subtracted central source is
11.1$\pm$1.1\,Jy. However, the method used here of separating the
photometry of the central component and the extended region is not
very accurate, because their flux was mixed in the observation process
by the PSF. In Sect.\,\ref{sec:decon} we present results of a more
precise deconvolution method. \\

\iprofile
\subsection{Comparison with previous Q-band imaging}
We compared our VISIR 18.72\,$\mu$m image with the SUBARU/COMICS
24.5\,$\mu$m image of \cite{2006ApJ...644L.133F}, applying the same
central component subtraction technique as described above. In
Fig.\,\ref{fig:contour} we overplotted the central component-subtracted 
COMICS image with white contours. It shows that the eastern
emission peak shifts to the south when going from 18.72 to
24.5\,$\mu$m. This cannot be explained with a disk inclination or with
flaring of the disk and is therefore an indication of azimuthal dust
opacity variation, which must be linked to an azimuthal density
variations. We will return to this point in Sect.~\ref{sec:outer}.

We also analyzed the mean surface brightness as a function of
distance from the star.  \ct{In Fig.\,\ref{fig:iprofile} we plot these spatial
profiles for our VISIR image and for the COMICS image. The position 
of the star was assumed to coincide
with the locus of the emission optimum, which was derived to an accuracy 
of a tenth of a VISIR pixel with a two dimensional Gaussian fit.} Following 
the same analysis approach as in
\cite{2006ApJ...644L.133F}, we treated the eastern and western
components separately: the eastern profile was averaged over the $PA$
= 0-180$^\circ$ and the western profile was averaged over the $PA$ =
180-360$^\circ$.  The observed
profiles in the top panel show that the flux ratio of $F_{\rm
18.72\,\mu m}/F_{\rm 24.5\,\mu m}$ substantially decreases from the
inner to the outer region. The bottom panel shows that the COMICS
image is more extended. These features are consistent with a radially
decreasing dust temperature.

\decon
\deconpro
\azprofile
\subsection{Deconvolution}
\label{sec:decon}
A more precise method to separate the central and outer emission
components is to deconvolve the image. We deconvolved the Q2
image with the MEM-multi-resolution method
(\citealt{1996A&AS..118..575P}) and obtained the image shown in
Fig.\,\ref{fig:decon}. The spatial resolution in the deconvolved image
was limited to 0.1$\arcsec$, which is on the order of the Shannon
limit imposed by the spatial sampling. At that resolution the image is
composed of two components. The central component is marginally
resolved (FWHM of 0.15$\arcsec$) and has an integrated flux of 10.84
Jy. The second component, located farther away at a distance of
$\sim$0.8$\arcsec$, has a narrow ring-like shape. The ridge drawn by
the maximum of emission in the ring can be fitted by an ellipse with
an aspect ratio of 1.2 ($i$ = 20$^{\circ}\pm$10$^\circ$ from face-on)
and a position angle of $PA$ = 120$^\circ\pm$20$^\circ$. These
numbers are slightly different from those obtained from the central
component-subtracted image. We will return to this in
Sect.\,\ref{sec:comp}. The emission ring has a mean distance to the
star (measured on the semi-major axis) of 0.8$\arcsec$ (130\,AU) and a
FWHM of 0.2$\arcsec$ (30\,AU, see Fig.\,\ref{fig:deconpro}).
Figure\,\ref{fig:deconpro} displays the mean radial profile of the
deconvolved image, obtained by averaging the emission over
360$^{\circ}$. The surface brightness as a function of azimuth
averaged over the radii from 0.5$\arcsec$ to 1.1$\arcsec$ has minima
of 0.6\,Jy/arcsec$^2$ at $PA = 0^\circ$, 180$^\circ$, and 230$^\circ$
and a maxima of 2.4\,Jy/arcsec$^2$ at $PA$ = $\sim$60$^\circ$ and
1.7\,Jy/arcsec$^2$ at $PA$ = $\sim$300$^\circ$(see
Fig.\,\ref{fig:az_profile}).

\section{VISIR spectroscopy analysis}
\label{sec:vspcana}
\VISIRSPC
\subsection{N-band}
To achieve a continuous VISIR \ch{low-resolution} N-band \ch{spectrum},
we spliced the four wavelengths settings together. We scaled them with 
the ratio of the median values of the overlapping
regions. The flux of the spectrum was then scaled to the flux of the
Spitzer spectrum at 10.6\,$\mu$m with an average factor of 1.24. We present 
the resulting spectrum in Fig.\,\ref{fig:VISIRSPC}. We also
plotted the Spitzer and TIMMI2 spectra for comparison. \ch{The excess 
at the blue end of the Spitzer spectrum is suggestive of extended low 
surface brightness PAH emission. Although there is no direct 
observational evidence in our data, this extended emission could be linked 
to the components visible at 18.72 and 24.5\,$\mu$m that largely fall outside 
the VISIR and TIMMI2 slits (see Fig.\,\ref{fig:slitpos}). The small 
differences at 9.2, 10.0 and at 12.6\,$\mu$m are not well understood.}

\QSPC
\profile
\subsection{Q-band}
Figure\,\ref{fig:slitpos} shows the chosen slit orientation of the
VISIR medium-resolution Q-band spectrum. The spectrum contains a
contribution from the central component and a contribution from the
eastern component. In Fig.\,\ref{fig:QSPC} we plot the integrated
spectrum. The spectral setting centered on 18.18\,$\mu$m has been
scaled by a factor 1.15 to continuously connect to the spectral
setting centered on 17.82\,$\mu$m. The wavelength regions where the
atmospheric transmission as measured with the standard star is less
than 30\% of the maximum are masked out with yellow bands. We
overplotted in thick green the Spitzer spectrum, which shows a
significant difference in slope compared with the VISIR spectrum.  The
bluer slope of the VISIR spectrum is in accordance with expectations,
because the VISIR slit has selected the inner, hotter parts of the disk,
while the Spitzer spectrum covers the entire disk, including all
colder outer parts.

The chosen slit orientation allows us to spatially separate the
spectra of the central and eastern components. This is illustrated in
Fig.\,\ref{fig:profile}, where we plotted the spatial profile of
the spectral section centered on 17.82\,$\mu$m after a collapse of the
wavelength dimension. In green we overplotted the calibrator. The
eastern (left) and western (right) components are both significantly
detected. Even the central component seems to be marginally resolved,
although the statistics are poor. We also extracted from the 2D spectral 
image two five pixel wide traces
centered on the central emission and the eastern peak respectively.
Their ratio showed no reliable spectral structures.

\section{Modeling}
\label{sec:model}
\ch{To better understand the circumstellar matter around 
HD\,142527 we created a detailed geometric model that is capable of 
reproducing the described observations. We begin by explaining the 
principles of the model, the basic assumptions, and the setup. Then we constrain 
the free parameters and construct the final best-fit model using radiative transfer 
calculations and the observations. This final model is tuned to accurately fit the
SED, the Spitzer and ISO IR spectra, and the general features of the IR images 
of the star. We conclude our modeling effort with a discussion of the 
discrepancies between the simulated and observed data.}

\ch{\subsection{Model basics and setup}
Our model consists of an axisymmetric structure that surrounds a single
star. This structure is defined by its density distribution, composition, 
inclination, position angle, and inner and outer radius. We use 
the Monte Carlo radiative transfer code MCMax 
\citep{2009A&A...497..155M} to obtain a corresponding temperature distribution
and simulated observables. To constrain the model parameters
we start with a basic setup which is refined after the radiative transfer 
calculation and comparison of simulated data with observables. In the
radiative transfer calculation we only take the flux from the
central star into account. Possible heating by viscous processes in the disk
can be ignored because the mass accretion rate of the disk is negligible.}

\ch{For the basic setup of the surface density of the disk we use a
simple power law}
\begin{equation}
\Sigma(r) = \Sigma_{\rm 0} \cdot \left(\frac{r}{r_{\rm 0}}\right)^{p},
\end{equation}
\ch{where $r$ denotes the distance to the central star, the exponent $p$
is set to the commonly used value -1 (see e.g. 
\citealt{2006ApJ...640L..67D}), and the scaling factor, $\Sigma_{\rm 0}$, 
determines the mass of the disk. It is clear from the images 
(Figs.~\ref{fig:contour} and \ref{fig:decon}) that the surface density of the disk is not
continuous. We therefore split the disk into an inner and an outer disk
separated by a gap. The structure of the two regions is treated 
separately.}

\ch{The vertical density distribution of the disk is solved by iterating
the density and temperature structures until they are
self-consistent under the assumption of hydrostatic equilibrium
(see e.g. \citealt{2007prpl.conf..555D}). Very small grains
couple well to the gas and thus follow this hydrostatic equilibrium
density distribution. However, larger grains may partly decouple
from the gas and settle towards the midplane of the disk through
gravitation. To simulate this we added an extra parameter that scales
the height of the dust in the disk ($\Psi$; see
\citealt{2007A&A...471..173R}; or \citealt{2009A&A...502L..17A}). For values of
$\Psi$ lower than unity, this $\Psi$ can be interpreted as a settling
of the grains to the midplane, hence we refer to this as the
sedimentation parameter.}

\ch{We fixed the composition and the grain size and shape distribution 
of the silicate component of the dust of the inner disk to
be equal to that obtained by \citet{2005A&A...437..189V}. In that
paper the continuum flux was represented by emission from a single
blackbody, which is sufficient for modeling the 10\,$\mu$m feature. 
Here, we use a continuum opacity in the form of amorphous
carbon grains with a variable abundance to improve the fitting 
accuracy in the optical and millimeter regimes. For the shape of the
carbon grains, we used the distribution of hollow spheres (DHS; see
\citealt{2005A&A...432..909M}), and for the refractive index we used
the data by \citet{1993A&A...279..577P}. The continuum component 
could also be reproduced by small grains of metallic iron and/or iron sulfide, 
as well as large grains of various dust species. However, this will have no great 
influence on the modeled geometry. For the outer disk we used the same
basic compositional setup except for the addition of small water-ice grains.
This is motivated by the clear water-ice feature around 45\,$\mu$m 
as seen in the ISO spectrum.} 

\ch{We assume that all types of dust species in the disk have the same 
temperature as the gas. This can be taken to mean that the different 
species are in thermal contact. To simulate the presence of 
composite grain-like ice-mantles on silicate and carbon grains we simply 
added the opacities of the individual species. It would of course be physically more 
realistic to take the interactions of the different components in the composite grain into
account \citep[see e.g.][]{2008A&A...489..135M}, which would
likely lead to a slightly different best-fit composition and grain
size distribution. However, for understanding the geometrical
characteristics and the general composition of the disk, the simple
approach suffices.}

\ch{The prime puzzle of HD\,142527 is its extremely large flux in the
far-IR. The wavelength-integrated flux in the IR regime is almost equal
to the integrated flux from the star ($F_{\rm IR} = 0.92 \cdot
F_*$). Any successful disk model will first and foremost need to
explain this remarkable observation.  Secondly, the peculiar
spatial distribution of the IR emission as seen at 18.72 and
24.5\,$\mu$m will need to be reproduced and the mid-IR
interferometric visibilities have to be matched as well. Below we discuss
several options to correctly model the SED and show how we arrived at
our final best-fit model using the IR images and mid-IR
interferometric constraints.}

\miprofile
\modelima
\mspectra
\ch{\subsection{Structure of the outer disk}
\label{sec:fitting}
The central ingredient to obtain a high $F_{\rm IR}/F_*$ ratio is to
move the bulk of the disk mass a large radial distance from the star where it
can form a very high vertical wall and reprocess a large fraction of the stellar flux
(see e.g. \citealt{Deroo2007}). This can be understood in the
following way. The fraction of the solid angle covered by the disk is
approximately $H/r$, where $H$ is the height of the optically thick
disk over the midplane and $r$ is the distance from the star.  $H$ can
be written as the pressure scale height $h_p$ times an approximate 
constant (\citealt{1997ApJ...490..368C}).  The standard disk theory shows that 
$h_p=c_s/\Omega_{\rm K}$ where $c_s=\sqrt{kT/(\mu m_p)}$ is the isothermal 
sound speed at temperature $T$, with $\mu m_p$ being the mean molecular 
weight.  $\Omega_{\rm K}=\sqrt{GM_{\star}/r^3}$ is the local Kepler frequency.  
The temperature of a freely irradiated wall decreases as a function of
distance with $T_{\rm wall}\propto r^{-1/2}$, so $c_s\propto r^{-1/4}$
and $H/r \propto r^{1/4}$.  This means that the larger the distance from the star, the higher
$H/r$ and $F_{\rm IR}/F_*$ will be. } 

\ch{We find that a gap
extending to 130 AU will allow the disk wall to rise high enough to 
account for most of the IR excess. The inclination and position angle of the disk were 
constrained by fitting the IR images and their radial surface brightness profiles 
(Figs.~\ref{fig:innerdisk}, \ref{fig:contour}, \ref{fig:decon}, \ref{fig:deconpro}, 
and \ref{fig:miprofile}). The arc-like structures are well reproduced by the radiation
from the inner edge of an inclined outer disk (see Fig.~\ref{fig:modelima}). The 
outer radius was constrained by fine-tuning the fit to the narrow bump around 
40\,$\mu$m in the SED (see Fig.~\ref{fig:mspectra}). The disk mass was constrained 
by fine-tuning the fit of the flux level of the SED fit in the millimeter regimes. We note 
that for the necessary scale height of the outer disk to be achieved in a natural way, it is
important that the stellar radiation reaches it largely unhindered.  As we will see, 
combined with other observations, this poses significant constraints on the structure 
of the dust close to the star.}

\modelfit 
\MIDI 
\ch{\subsection{Structure of the inner disk}
The inner radius is constrained by fine-tuning the SED fit in the 
near-IR regime. For the vertical distribution 
of the material close to the star we consider two scenarios. In the first the 
material is confined in an optically thick disk and in the second the material
is confined in an optically thin dust component with very large scale height.} 

\ch{\emph{An optically thick inner disk?} When we include a standard 
inner disk in the radiative transfer, the
disk is strongly optically thick and casts a shadow on the outer disk.
Furthermore, we need to artificially increase the scale height of this disk by 
roughly a factor of 3.  While this produces the
correct near-IR flux, it also casts a wide shadow onto the
\emph{outer disk}. The resulting loss of irradiated outer disk surface, and 
consequently the decrease in temperature, decreases the
scale height and emission from the outer disk.  To preserve the match
with the observed mid- and far-IR flux, we would need to
artificially increase the scale height of the outer disk as well, and that by a factor of
$2.5$ over the self-consistent hydrostatic equilibrium value.
A model like this has several shortcomings. First of all, there is no
physically realistic explanation for the increased scale height
of the \ca{outer} disk over hydrostatic equilibrium, because this would require
\ca{very} high gas temperatures in the mid-plane, for which \ca{we have}
no justification.}  Second, the resulting model images are more
axisymmetric than the observed IR images. And third, 
we compared the mid-IR visibilities obtained by \citet{2004Natur.432..479V} with MIDI
with those obtained from this model and found that the model visibilities
are a factor 1.5 too high (see Fig.\,\ref{fig:MIDI}). This means that
the emission is too much peaked toward the center of the disk.

\ch{\emph{A halo-like component?} As suggested by 
\citet{2010A&A...512A..11M}, a way to avoid the extended
shadow on the outer disk is to put the material in the inner region
into an optically thin dust component, \ca{which extends to a very high altitude, for
example in a halo (\citealt{2003MNRAS.346.1151V,2006ApJ...636..348V})}. 
Although halos have been the topic of much controversy, there are also other stars that
show some indication of optically thin material \ca{extending to a high altitude}
in the inner disk (e.g. HD163209, \citealt{2010A&A...511A..74B}).  
We will discuss the possible origin of these halo-like structures in Sect.\,\ref{sec:inner}. 
For our radiative transfer simulations, we modeled this component as a spherical
halo, but we do not mean to imply that it has to be
fully spherical in reality.}  Using an inner dust shell we can obtain a
very good fit to both the images and the SED. Because we kept the
composition and grain size distribution of the halo the same as for the
inner disk, the halo causes a measure of gray extinction at optical
wavelengths ($\tau = 0.37$). This means a slightly higher stellar
luminosity $L$ = 20\,L$_\odot$ is required than would be the case
when only taking the interstellar extinction into account. The \ch{resulting} slight
decrease in scale height of the outer disk by a factor of 0.8 can be easily 
understood in terms of settling of the grains toward the midplane.  However, in order to
fit the spectral shape (i.e. the range of temperatures) of the
emission from the inner region, the powerlaw for the surface density
needs to be very flat ($\Sigma(r) \propto r^{-0.4}$). This leads to
mid-IR visibilities that are a factor 4 lower than the observed
visibilities (see Fig.\,\ref{fig:MIDI}).\\

\subsection{Final model}
\ch{We saw that the optically thick inner disk or the halo-like 
component alone do not deliver a satisfactorily model. The solution is to 
combine a low-mass inner disk in hydrostatic equilibrium with an optically 
thin halo-like component of dust.  This allows us to obtain a very good
agreement for the SED, the IR images and the mid-IR visibilities.}

\ch{We summarize our final best-fit model in Fig.\,\ref{fig:picto} and table\,\ref{tab:modelfit}. We
assume an axisymmetric disk that is slightly inclined. It has a
settled inner disk, which extends from 0.3 to 30\,AU, and a
sedimentation parameter of $\Psi = 0.4$. At 0.3\,AU the disk has a vertical 
height of $7 \cdot 10^{-3}$\,AU above the midplane (as traced by 
the radial $\tau$=1 surface in the optical). The dust reaches a temperature of 
$\sim$1600\,K, which we deem acceptable. There is a small mass in an
optically thin spherical halo between 0.3\,AU and 30\,AU. 
However, the outer radius of this halo is not
well constrained by the observations. The outer disk starts at
130\,AU. Vertically it was scaled to $\Psi = 0.9$ times the
hydrostatic equilibrium height, which puts the vertical height
of the disk wall at 60\,AU ($\tau$=1 surface).
The inner region contains silicate dust with the same
composition as obtained by \citet{2005A&A...437..189V}, and we used
37\% carbonaceous grains to represent all continuum opacity
sources. In order to correctly reproduce the wavelength position of the 
slope of the SED in the far-IR, we set the size of the grains in the outer 
disk to 2.5\,$\mu$m and used 20\% carbonaceous grains. We also added
45\% of crystalline water-ice to the dust-mixture of the outer disk 
to reproduce the observed 45\,$\mu$m ice feature.}

\spitzer
\subsection{Comparison of the model with the observations}
\label{sec:comp}
The disk model described above is capable to generally reproduce the
various observations. We are confident that the small discrepancies
between observational and modeling results reflect details in the
geometry and/or mineralogy that will not strongly affect the emerging
general picture of the system. Here we list the comparison and
discrepancies.

\emph{Photometry.} The fit to the literature photometry and ISO
spectrum is presented in Fig.\,\ref{fig:mspectra}. The flux level and
general slope of the photometry, as well as the features of the ISO
spectrum are reproduced. The model overestimates the long
wavelength part of the ISO spectrum by about 20\%. 

\emph{Spitzer.} The fit to the Spitzer spectrum is presented in
Fig.\,\ref{fig:spitzer}. The flux level and general shape are
reproduced, but the 10\,$\mu$m feature is overestimated by about 
10\%.  The forsterite features at longer wavelengths are
underestimated because we did not include \ch{small crystalline grains
in the outer disk. }
 
\emph{MIDI.} In Fig.\,\ref{fig:MIDI} the MIDI visibilities were
compared with the different models. Our final model reproduces the
spatial scales probed by these MIDI visibilities. The differences
are most likely caused by gradients in the crystallinity of the disk,
which we did not take into account.

\emph{VISIR and COMICS.} The modeled VISIR 18.72\,$\mu$m and COMICS
24.5\,$\mu$m images are compared with the observations in
Fig.\,\ref{fig:modelima}. The \ch{simulated} images were obtained by taking the
output of the model at the specified wavelengths and by convolving
them with their associated PSF. The overall spatial distribution of
the emission matches very well. In Fig.\,\ref{fig:innerdisk} we zoom
in on the radial profile of the inner disk as seen in the VISIR image.
In Fig.\,\ref{fig:miprofile} we plot the radial profile of the
mean surface brightness of the eastern and western sides of the
images. There is a slight discrepancy in the flux ratio of the
eastern lobe over the central component of the COMICS image.

\emph{VISIR deconvolved.} Figure\,\ref{fig:deconpro} shows the radial
profile of the mean surface brightness of the deconvolved VISIR image.
To model the deconvolved image we convolved the model output at
18.72\,$\mu$m with a 2D Gaussian with a FWHM of
0.1$\arcsec$. This way we mimicked the obtained resolution of the
deconvolved image. The fit is quite excellent, although the emission
in the gap is not reproduced. This likely indicates that some dust is
present in the gap. Figure\,\ref{fig:az_profile} shows the azimuthal
profile of the outer disk that results from the deconvolved image. The
$PA$ and the brightness of the eastern lobe are quite well reproduced,
but the brightness peak at the western side of the outer disk is not
reproduced. Furthermore, the image analysis gives a larger PA of
the disk itself then the model does. These are indications that the
density distribution and geometry of the outer disk are more complex
than assumed by our model (compare with Fig.\,\ref{fig:miprofile}), we
will return to this in Sect.\,\ref{sec:outer}.

\picto
\section{Discussion}
\label{sec:disc}
\subsection{General picture}
\ch{Our modeling effort shows that} the circumstellar matter around 
HD\,142527 consists of three geometrically distinct components, as depicted in
Fig.\,\ref{fig:picto}. First, there is a small inner disk structure
running from 0.3 to 30\,AU, with a puffed up inner rim and a
self-shadowed disk further out. Second, there is a small amount of
dust in an optically thin \ch{component \ca{extending to extremely high altitude}.} 
The third component is
a fairly massive outer disk, consisting of \ch{an extremely high}
wall at 130\,AU, which is in hydrostatic equilibrium, and a
self-shadowed outer part running up to 200\,AU. 

\ch{\subsection{Inner disk and halo-like structure}
\label{sec:inner}
In our model the inner disk and the halo-like structure together
are responsible for the near-IR flux, the slightly resolved inner
component of the VISIR image (see Fig.\,\ref{fig:innerdisk}),
most of the 10\,$\mu$m feature and the MIDI visibilities.} The inner
disk contains very little mass in small dust grains ($M_{\rm dust} =
2.5\cdot10^{-9}$\,M$_\odot$).  \ca{One likely explanation is that} all the material is already
locked up in planetesimals or larger bodies. The degree of
sedimentation of the dust compared with the gas ($\Psi$ = 0.4) seems to
point at an advanced state of evolution as well. \ch{Expressed in the radial
$\tau$=1 surface in the optical, the height of the material at 0.3\,AU reaches up to 
$7\cdot10^{-3}$\,AU.}

For the halo-like structure we estimated a dust mass of $1.3 \cdot
10^{-10}$\,M$_\odot$, which is one order of magnitude less than the inner
disk. However, the halo-dust particles dominate the near-IR flux. \ch{The
origin of the halo-like structure is not clear. It is generally expected that a 
dust halo decays naturally from stellar radiation pressure and gravity, therefore a
mechanism is needed to replenish it. A possible replenishment
scenario can be constructed with a dynamically highly excited
debris disk. An attempt to model the effects of such a scenario is underway (Krijt \& Dominik 2010, in
prep.). An observational hint for this type of replenishment was already
found by \cite{1998A&A...331..211M}, who observed variable
obscuration. There is also} the possibility that some dust is
entrained in the upper disk layers where gas and dust temperatures
strongly differ (\citealt{2009A&A...501..383W}), and the vertical
distribution of the gas is much more extended than in models like
ours, where gas and dust are coupled everywhere
(\ch{\citealt{2010arXiv1009.4374T}).  A third possibility may be a disk-wind
carrying small dust grains.}

\ch{A side effect of the presence of the halo-like structure is the
gray extinction at optical wavelengths and a resulting higher
estimate of the stellar luminosity as mentioned in
Sect.\,\ref{sec:parameters}. The values of the stellar radius, mass,
and age given in table\,\ref{tab:parameters} were also derived from
this new luminosity.}

\subsection{Disk gap}
Our analysis of the VISIR 18.72\,$\mu$m image and the modeling effort
(see Fig.\,\ref{fig:deconpro}) demonstrate the presence of an extensive \ch{disk}
gap ranging from 30 to 130\,AU, which is relatively void of small dust
grains. One might speculate about the presence of \ch{one or more} 
large Jupiter-mass planets in the gap that are responsible
for the clearing of the material. In the light of recent findings this
seems a very likely possibility.  \cite{2008Sci...322.1348M} and
\cite{2008Sci...322.1345K} found Jupiter-like planets in wide orbits
around young A-type stars with direct imaging.
\cite{2009A&A...493L..21L} consider the presence of a Jupiter-like planet
in the debris disk of $\beta$ Pictoris very
likely. Furthermore, models of the solar nebula generally rely on an
early formation of Jupiter, which then promotes the formation of the
terrestrial planets (\citealt{1996Icar..124...62P}). This argument
favors the presence of \ch{planetesimals or larger bodies} in the
inner disk (see Sect.\,\ref{sec:inner}).  

To check whether photoevaporation could be responsible for the
clearing of the gap we consulted \cite{2009ApJ...690.1539G}, who
describe the following scenario for the dispersal of a disk around a
1\,M$_\odot$ PMS star. In the outer regions of the disk FUV
photoevaporation dominates accretion and the disk is expected to
rapidly shrink to a truncation radius, $r_t \sim$150\,AU.  Under
favorable conditions a gap is created at 3-30\,AU, after which the
inner disk rapidly disappears on a viscous timescale of
$\lesssim10^5$\,yr. The outer disk is subsequently rapidly eroded
($\lesssim10^5$\,yr) from the inside out. The authors estimate disk
lifetimes of $\sim$10$^6$\,yr.  Because the gap of HD\,142527 is located
at larger radii and there is still material in the inner disk, it is
unlikely that photoevaporation is the dominant cause of the disk
geometry.

\subsection{Outer disk}
\label{sec:outer}
The most distinct and curious aspect of the appearance of HD\,142527
is its large IR excess ($F_{\rm IR} = 0.92 \cdot F_*$), which was
noted by \cite{2003A&A...398..607D}. We managed to self-consistently
explain this flux level by invoking the presence of \ch{an extremely
high} wall of material at 130\,AU which is in hydrostatic
equilibrium.  We found that the height of this material reaches up to
60\,AU, as expressed in the radial $\tau$=1 surface in the optical. This means that the
outer disk covers 42\% of the sky as seen from the star. It is this
covering fraction that allows the disk to reprocess so much stellar
light and emit IR radiation in \ch{two} relatively small solid angles
(\ch{2x29\%}). This wall is also the cause of the bright spot in the IR
images. A frontal irradiation over almost the entire vertical surface
is required in order for this wall to puff up the way it does. This is
guaranteed by the small solid angle of the flattened inner disk \ch{(2\%)}.

Further out the disk is self-shadowed and runs up to 200\,AU. From our
model we estimated the total dust mass in small grains in the outer
disk to be $1.0\cdot10^{-3}$\,M$_\odot$, consistent with our \ch{simple} analysis
of the millimeter data (see Sec.\,\ref{sec:anamilli}). Assuming a
gas-to-dust ratio of 100, the total disk mass is well below the
gravitational instability limits (see e.g.
\citealt{2001ApJ...553..174G}), but it is at the high end of the range
in disk masses (10$^{-3}$ to 10$^{-1}$\,M$_{\odot}$) found for other
Herbig stars by \cite{2004A&A...422..621A}.

The dust of the outer disk is also responsible for the scattering seen
in the Subaru images of \cite{2006ApJ...636L.153F}. \ch{The larger outer
radius stated by these authors is likely caused in part by PSF smearing and in part 
by a very low-density dust component. This component does not 
contribute to our observations, so we did not take it into account.} 
The cold dust of the outer disk is very processed, because it contains both 
crystalline $\element{H}_{2}
\element{O}$ ice (see also \citealt{1999A&A...345..181M}) and
forsterite. \cite{2008Ap&SS.313..101O} find convincing evidence for
strong gas depletion on the basis of \element{^{12}CO} (3-2) emission,
which is also an indication of advanced disk evolution.

The overall spatial brightness distribution was fairly well modeled,
but from the overlay of the images at 18.72 and 24.5\,$\mu$m in
Fig.\,\ref{fig:contour} it is clear that the peak of the emission of
the eastern lobe shifts to the south when going to longer
wavelengths. This asymmetry cannot be explained with an
axisymmetric density distribution; thus this is an indication of
azimuthal variation in the density. Also, Figs.\,\ref{fig:decon} and
\ref{fig:modelima} show that there are some discrepancies between the
observed images and the modeled images in the azimuthal
direction. Previous modeling efforts of protoplanetary disks \ch{have} made
the existence of spiral density waves likely (see
\citealt{2007prpl.conf..607D} for a review). It is possible that
we are witnessing the signature of these waves here in HD\,142527.

To add to the complexity of the geometry, the sub-millimeter mapping by
\cite{2008Ap&SS.313..101O} showed that the spatial distribution of the
bulk of the material is in an arc-like structure enclosing the central
star. Their image shows concentrations to the northeast and
northwest and a surprising absence of material to the south. The
similarity to our deconvolved 18.72\,$\mu$m is striking (see
Fig.\,\ref{fig:decon}). From our azimuthal analysis we concluded that
the emission maxima are actually at $PA=60^\circ$ and $PA=-60^\circ$ 
(see Fig.\,\ref{fig:az_profile}). The resemblance that
comes to mind is that of the Trojan asteroids of Jupiter, who reside
in the two Lagrangian points of stability that lie 60$^\circ$ ahead
and behind the planet's orbit. This would imply the presence of a large
planetary body 0.8$\arcsec$ (120\,AU) north of the star. Such a body
could also cause the inner edge of the outer disk to have an eccentric
shape. This could explain the discrepancy found in the PA of the disk
between the modeling and the image analysis
(Sect.\,\ref{sec:decon}). A modeling of the gravitational dynamics is
called for to determine whether this configuration is physically
possible and to constrain the exact orbital parameters and mass of
such a body.

\section{Conclusions}
\label{sec:conclusions}
We presented an observational and modeling study of the
circumstellar environment of HD\,142527. The main conclusions of our
work can be summarized as follows:

\begin{itemize}

\item \ct{We obtained observations with SEST and ATCA, which pin the
millimeter flux to the optical position of HD\,142527. From the spectral 
slope in the millimeter we derived an opacity index of 
$\beta$ = 1.0$\pm$0.1, indicative of a relatively small grain size. From the 
SEST flux at 1.2\,mm we derived a total dust mass in small grains of 
1.0$\cdot$10$^{-3}$\,M$_{\odot}$.}

\item Mid-infrared imaging of the disk surrounding HD\,142527 using
VISIR confirms the presence of a large gap extending from $\sim$30 to
$\sim$130 AU. We find that the disk is unresolved in the 10\,$\mu$m window
with no sign of emission from the outer disk.  The emission in the
VISIR 18.72 $\mu$m image is dominated by the marginally resolved inner
disk and by the inner rim of the outer disk.  A comparison between
VISIR 18.72 $\mu$m and SUBARU 24.5 $\mu$m imaging
\citep{2006ApJ...644L.133F} shows good agreement; the spatial scale of
the emission increases with wavelength, which is consistent with an outward
decreasing dust temperature.

\item We modeled the VISIR, SUBARU, MIDI, Spitzer, ISO and millimeter
data with the MCMax radiative transfer code
\citep{2009A&A...497..155M}. The large infrared excess can be 
understood with a \ct{self-consistent disk model with the following properties: (i) a 
geometrically flat inner disk running from 0.3\,AU up to 30\,AU, (ii) an 
optically thin halo-like component of dust in the inner disk regions, (iii) an 
extensive disk gap, and (iv) a massive outer disk running from 130\,AU up to 200\,AU. 
The inner rim of the outer disk is directly irradiated by the
central star, which causes a huge scale-height (h = 60\,AU). The derived mass in 
small dust grains is consistent with our millimeter analysis 
($1.0\cdot10^{-3}$\,M$_{\odot}$). }

\item \ct{As a consequence of the gray extinction caused by the halo-like component 
we obtained new stellar parameters, including a stellar luminosity of L = 20$\pm$2\,L$_{\odot}$ 
and a stellar age of log t = 6.7$\pm$0.4\,yr.}

\item The deconvolved VISIR image shows azimuthal intensity variations
that suggest a more complex structure of the outer disk than can be
accounted for with our model. This suggests a non-axisymmetric 
distribution of material.

\item \ct{The presence of the halo-like component, the low mass in small 
grains and the high degree of sedimentation in the inner
disk suggest the presence of planetesimals or larger bodies in the inner
disk. The large disk gap, the highly processed nature of the grain population
and the large mass of the outer disk suggest that planet
formation is also on-going in the outer disk. HD\,142527 is thus likely to evolve 
into} a system similar to those recently found around young A-type stars 
\citep{2008Sci...322.1348M, 2008Sci...322.1345K}. 

\end{itemize}

\begin{acknowledgements}
This research was sponsored by NWO under grant number 614.000.411.
Part of this work was supported by the German \emph{Deut\-sche
For\-schungs\-ge\-mein\-schaft, DFG\/} project number Ts~17/2--1.
M. Min acknowledges financial support from the Netherlands
Organization for Scientific Research (NWO) through a Veni grant.
E. Pantin acknowledges financial support from the Agence Nationale
de la Recherche (ANR) of France through contract
ANR-07-BLAN-0221. We gratefully acknowledge \ch{helpful} discussions
with A. Quillen.
\end{acknowledgements}

\bibliographystyle{aa}
\bibliography{../biblio}

\end{document}